\documentclass[pra,aps,twocolumn,floatfix,superscriptaddress]{revtex4-1}
\usepackage{amsfonts}
\usepackage{amsmath}
\usepackage{tikz}
\usepackage{graphicx}
\usepackage{amssymb}
\usepackage{mathrsfs}
\usepackage{hyperref}
\hypersetup{colorlinks=true, linkcolor=blue, citecolor=blue, 
urlcolor=blue}
\newtheorem{theorem}{Theorem}
\newtheorem{definition}{Definition}
\newtheorem{prooflem}{Proof of Lemma}
\newtheorem{proofthm}{Proof of Theorem}
\newtheorem{proofcor}{Proof of Corollary}
\newtheorem{lemma}{Lemma}
\newtheorem{cor}{Corollary}
\newcommand{\eqdef}{\ensuremath{\stackrel{.}{=}}}
\newcommand{\ket}[1]{\ensuremath{|{#1}\rangle}}
\newcommand{\bra}[1]{\ensuremath{\langle#1|}}
\newcommand{\braket}[2]{\ensuremath{\left\langle{#1\vphantom{#2}}\right.\!\left|\vphantom{#1}#2\right\rangle}}
\newcommand{\ketbra}[2]{\ensuremath{\ket{#1}\!\bra{#2}}}
\newcommand{\idxof}[1]{\ensuremath{\tilde{#1}}}
\newcommand{\ncell}{\ensuremath{m}}
\newcommand{\nvert}{\ensuremath{n}}
\newcommand{\zerovec}{\ensuremath{\mathbf{0}}}
\newcommand{\lblblk}{\ensuremath{a}}
\newcommand{\lblvec}{\ensuremath{b}}

\begin{document}

\title{Perfect quantum state transfer of hard-core bosons on weighted path 
graphs}
\author{Steven J. Large}
\affiliation{Department of Physics, University of Guelph,
Guelph, Ontario, Canada N1G 2W1}
\author{Michael S. Underwood}
\affiliation{Centrality Data Science, Calgary, Alberta, Canada T2E 0K6}
\author{David L. Feder}
\email[Corresponding author: ]{dfeder@ucalgary.ca}
\affiliation{Institute for Quantum Science and Technology, and
Department of Physics and Astronomy,
University of Calgary, Calgary, Alberta, Canada T2N 1N4}
\date{\today}

\begin{abstract}
The ability to accurately transfer quantum information through networks is an 
important primitive in distributed quantum systems. While perfect quantum state 
transfer (PST) can be effected by a single particle undergoing continuous-time 
quantum walks on a variety of graphs, it is not known if PST persists for
many particles in the presence of interactions. We show that if single-particle 
PST occurs on one-dimensional weighted path graphs, then systems of hard-core 
bosons undergoing quantum walks on these paths also undergo PST. The analysis 
extends the Tonks-Girardeau ansatz to weighted graphs using techniques in 
algebraic graph theory. The results suggest that hard-core bosons do not 
generically undergo PST, even on graphs which exhibit single-particle PST.
\end{abstract}


\maketitle

\section{Introduction}

Perfect quantum state transfer (PST) was first conceived in the context of
coupled spin networks~\cite{Christandl2004,Christandl2005}, in which quantum 
information encoded in a given spin (qubit) is transferred with unit fidelity 
to a different spin elsewhere in the network. PST was shown to occur between 
spins across the diameter of hypercubes, as well as on the one-dimensional (1D) 
spin chains whose spatially varying (but time-independent) coupling constants 
derive from a suitable projection of the hypercube. A complete set of coupling 
constants for 1D chains has been subsequently characterized~\cite{Wang2011}.
The single-excitation subspace of a spin network is equivalent to a single 
particle undergoing a continuous-time quantum walk on a graph with the same 
connectivity~\cite{Hines2007}, and in this context PST has been shown to occur 
on variety of graphs including graph quotients and joins, circulants, 
double cones, cubelike graphs, and signed graphs, among others~\cite{Saxena2007,Facer2008,Bernasconi2008,Jafarizadeh2008,Angeles-Canul2009,Basic2009a,Basic2009b,PST2,Ge2011,PST3,Brown2013}. 
If local control is permitted at the endpoints, then PST is possible on a 
wider assortment of networks~\cite{Burgarth2005,Burgarth2006}.
PST is a basic primitive in quantum communication across quantum networks, and 
can be a building block in the construction of approaches to universal quantum
computation~\cite{Kay2010}. References \cite{Godsil-main} and \cite{Kay2011} provide 
relatively recent reviews of PST on graphs and spin networks, respectively. 

Any physical implementation of PST within the framework of continuous-time
quantum walks on graphs will likely involve multiple particles, for example 
ultracold bosons in optical lattices~\cite{Bloch2005,Lewenstein2007,Bloch2012}. 
Non-interacting bosons undergo PST if the single-particle graph also supports
PST~\cite{Yung2005}; many-boson systems also yield a hierarchy of (generally
weighted) graphs that enable single-particle PST~\cite{Feder2006,PST3}. 
Physical particles generically interact with each another; for example, 
ultracold atoms in optical lattices are well-described by having effective 
on-site interactions. Yet few results are known about the occurence of PST for 
interacting bosons, beyond an early analysis within the context of the 
Bose-Hubbard model~\cite{Wu2009} or for very small graphs~\cite{Underwood2012}.

In the limit of very strong repulsive interactions between bosons, known as the
hard-core limit, at most one boson may occupy a given lattice site or graph 
vertex~\cite{Cazalilla2011}. The restriction on site occupancy for hard-core 
bosons is reminiscent of the Pauli exclusion principle for (spinless or 
spin-polarized) fermions. Indeed, for one-dimensional lattices the ground state 
can be obtained exactly using the Tonks-Girardeau (TG) 
construction~\cite{TG,TG2}. The TG eigenstates correspond to those of 
non-interacting fermions which are then expressly symmetrized by the 
application of a `unit antisymmetry operator' ${\mathcal A}$ to ensure that the
total bosonic wavefunction is symmetric under particle exchange. For the ground
state (which is necessarily all-positive for bosons), this is equivalent to 
taking the absolute value of the many-particle wavefunction, which is 
non-analytic. The TG state has been realized experimentally in an ultracold 
bosonic gas~\cite{Paredes2004}.

In fact, the XY model generally used to study PST on 1D spin chains maps 
exactly via the Jordan-Wigner transformation~\cite{JW} to a model of 
non-interacting fermions, with the variable coupling constants mapping to the 
variable hopping amplitudes. Thus, truly non-interacting fermions on 1D 
lattices (path graphs) that exhibit PST should also undergo PST. That said,
the presence of the unit antisymmetry operator that maps all fermionic 
eigenstates to those for hard-core bosons in the TG construction could have 
important consequences for PST, whose existence depends crucially on 
constructive phase interference. Likewise, the presence of the ${\mathcal A}$ 
complicates the calculation of other expectation values for hard-core bosons, 
compared to those for non-interacting fermions, such as the single-particle
correlation function~\cite{Cazalilla2011}. Thus, in principle hard-core bosons 
on path graphs could display different dynamics from those of non-interacting 
fermions on the same graphs. 

In this work, we prove that hard-core bosons on path graphs with edge weights 
derived from hypercubes indeed undergo PST. The proofs make heavy use of 
algebraic graph theory as well as the TG construction for {\it distinguishable}
bosons framed in a graph theory context, which is shown to remain exact for 
weighted paths. In order to subsequently map the system to 
identical bosons, we construct an extension of simple graph equitable 
partitioning to general weighted graphs. Because hard-core bosons are not able 
to pass through one another as they undergo PST, the bosons must retain their 
original ordering throughout their evolution. The symmetry associated with PST 
in these systems is found to be a combination of parity and particle 
re-ordering, and the existence of PST hinges on the fact that for path graphs
the associated symmetry operator commutes with the unit antisymmetry operator 
that maps non-interacting fermions to hard-core bosons. The inference is that 
hard-core bosons on more general graphs are unlikely to exhibit PST, even when 
the underlying graphs exhibit PST for single particles.

The manuscript is organized as follows. Section~\ref{sec:background} reviews 
the essential material that will be needed in the discussion of PST for 
hard-core bosons on weighted path graphs. This includes the properties of 
continuous-time quantum walks on graphs, equitable partitioning, PST on graphs, 
the theory of many-boson systems on paths (both non-interacting and hard-core 
bosons), and the Tonks-Girardeau construction. The main results of the work are 
presented in Section~\ref{sec:results}; the characteristics of the Cartesian 
power graphs under vertex deletions (due to multiple bosonic occupation of a 
given vertex) are given, followed by the description of the important 
symmetries of the system. These ingredients are used to prove the main theorem. 
We conclude in Section~\ref{sec:discussion} with a discussion of the 
ramifications of these results for more general graphs.

\section{Background}
\label{sec:background}

\subsection{Continuous-time quantum walk on graphs}
\label{sec:backwalk}

Consider finite, connected, and undirected graphs ${\mathcal G}=(V,E,w)$ with 
vertex set $V$ and edge set $E\subseteq V\times V$ connecting the 
$n\eqdef|V|$ vertices. The weight function is $w:E\to\mathbb{R}^+$, so that the 
adjacency matrix $A_{\mathcal{G}}$ associated with the graph ${\mathcal G}$ has 
elements $A_{\mathcal{G}}=w(u,v)\eqdef w_{uv}$ where $(u,v)\in E$. Because the 
graph is undirected the adjacency matrix is symmetric, $w_{uv}=w_{vu}$. The 
time-evolution operator for a single particle undergoing a continuous-time 
quantum walk on a graph is defined as~\cite{Farhi1998}
\begin{equation}
U_{\mathcal{G}}(t)\eqdef e^{-itA_{\mathcal{G}}}.
\label{time-evolution}
\end{equation}
The adjacency matrix can be expressed in its spectral decomposition
\begin{equation}
A_{\mathcal{G}} = 
\sum\limits_{j=0}^{n-1}\tilde{\lambda}_j|z_j\rangle\langle z_j|\label{eq:spec},
\end{equation}
where $|z_j\rangle$ and $\tilde{\lambda}_j$ are its orthonormal eigenvectors 
and eigenvalues, respectively. The time-evolution 
operator~(\ref{time-evolution}) then becomes
\begin{equation}
U_{\mathcal{G}}(t) = \sum\limits_{j=0}^{n-1}e^{-it\tilde{\lambda}_j}
|z_j\rangle\langle z_j|.
\label{eq:UG}
\end{equation}
The time evolution of an initial state $|\psi(0)\rangle$ is effected by direct 
application of the time evolution operator, $|\psi(t)\rangle = U_{\mathcal{G}}
(t)|\psi(0)\rangle$.

One can rewrite Eq.~(\ref{eq:UG}) as
\begin{equation}
U_{\mathcal{G}}(t) = e^{-it\tilde{\lambda}_i}\sum\limits_{j=0}^{n-1}
e^{it(\tilde{\lambda}_i-\tilde{\lambda}_j)}|z_j\rangle\langle z_j|,
\label{eq:UG2}
\end{equation}
with $\tilde{\lambda}_i$ an arbitrary eigenvalue. It is easy to verify that if 
the eigenvalues satisfy the ratio condition
\begin{equation}
\frac{\tilde{\lambda}_i - \tilde{\lambda}_j}
{\tilde{\lambda}_k - \tilde{\lambda}_l}\in \mathbb{Q} 
,\label{eq:ratio}
\end{equation}
for all possible combinations of $\tilde{\lambda}_r$, $r\in\{i,j,k,l\}$ (except
$\tilde{\lambda}_k=\tilde{\lambda}_l$), then the graph is periodic. In
other words, there exists a time $t=\tau'$ where $U_{\mathcal{G}}(\tau')$ is
equivalent to an $n\times n$ identity matrix 
$I_n=\sum_j|z_j\rangle\langle z_j|$ up to an 
unimportant phase. Alternatively, defining the vertex state $|u\rangle$ as the 
unit vector corresponding to the vertex 
$u$, one can write $U_{\mathcal{G}}(\tau')|u\rangle=\gamma'|u\rangle$ for all 
vertex states $|u\rangle$, where $\gamma'$ is a complex number with unit norm, 
i.e.\ $|\gamma'| = 1$.

Perfect quantum state transfer (PST) is said to be achieved if there exists 
a time $\tau$ such that 
\begin{equation}
U_{\mathcal{G}}(\tau)|u\rangle = \gamma|v\rangle
\label{eq:PST}
\end{equation}
is satisfied for two distinct vertex states $|u\rangle$ and $|v\rangle$, where
again $|\gamma|=1$. If there exists PST between 
vertices $|u\rangle$ and $|v\rangle$ at time $t=\tau$, then there also exists 
PST from $|v\rangle$ to $|u\rangle$ at the same time. This implies that in 
twice the time, there is PST from $\ket{u}$ back to itself:
$U_{\mathcal{G}}(2\tau)|u\rangle = U_{\mathcal{G}}(\tau')|u\rangle 
= \gamma'|u\rangle$. Thus the presence of 
PST ensures vertex periodicity, though it is important to underline that the 
converse is not generally true~\cite{Godsil-main}. That said, if PST exists for 
a periodic graph, then it will occur in a time $\tau=\tau'/2$. The ratio 
condition~(\ref{eq:ratio}) is therefore a necessary but not sufficient 
condition for PST to take place on a graph $\mathcal{G}$.

The adjacency matrix for two distinguishable particles undergoing a quantum
walk on the graph ${\mathcal G}$ is given by the Cartesian product of the
graph adjacency matrix with itself, 
$\mathcal{G}\Box \mathcal{G}$~\cite{PST3}. For two general graphs
$\mathcal{G}$ and $\mathcal{H}$ on $n$ and $m$ vertices, respectively, the 
Cartesian product is defined in terms of their adjacency matrices as
\begin{equation}
A_{\mathcal{G}\Box \mathcal{H}} = A_\mathcal{G}\otimes I_m + 
I_n\otimes A_\mathcal{H},
\label{eq:CartesianProduct}
\end{equation}
where $\otimes$ 
denotes the tensor product. If both $\mathcal{G}$ and $\mathcal{H}$ are path
graphs $P_n$ and $P_m$, their Cartesian product is the $n\times m$ grid. The 
$k$-fold Cartesian power of the graph $\mathcal{G}$ on $n$ nodes is 
$\mathcal{G}^{\Box k} = \mathcal{G}\Box\mathcal{G}\Box\cdots\Box\mathcal{G}$, 
which is explicitly
\begin{eqnarray}
A_{\mathcal{G}^{\Box k}} &=& \left( A_{\mathcal{G}}\otimes I_n\otimes\cdots
\otimes I_n\right)+\left(I_n\otimes A_{\mathcal{G}}\otimes\cdots\otimes I_n
\right) \nonumber\\
&+&\cdots +\left( I_n\otimes\cdots\otimes I_n\otimes A_{\mathcal{G}}\right).
\label{eq:construct-config}
\end{eqnarray}
in terms of its adjacency matrix. Because all of the terms in the above sum
are mutually commuting, the time-evolution operator of the $k$-fold Cartesian 
power is
\begin{equation}
U_{\mathcal{G}^{\Box k}}(t) = \bigotimes\limits_{i=1}^{k}U_{\mathcal{G}}
(t),\label{time-separation}
\end{equation} 
the $k$-fold tensor product of the time-evolution operator on the original
graph $\mathcal{G}$. 

Eq.~(\ref{time-separation}) ensures that if there exists PST, or periodicity of 
any vertex, on the underlying graph, then the same property exists in the 
Cartesian power graph. Consider for example PST on the hypercube 
$\mathcal{Q}_n$, which is defined as the $n$-dimensional analog of 
the cube. It is constructed by a recursive product of the path graph $P_2$ with 
itself, so that
\begin{eqnarray}
	\mathcal{Q}_2 &\eqdef& P_2, \nonumber  \\
	\mathcal{Q}_{n>2} &\eqdef& P_2\Box \mathcal{Q}_{n-1}.
\end{eqnarray}
Because there exists PST on $P_2$ 
in time $t=\pi/2$, then there also must exist antipodal PST on the hypercube 
$\mathcal{Q}_n$ in the same time.

\subsection{Equitable partitioning}
Equitable partitioning is a powerful tool in the analysis of graphs, in which 
vertices are grouped together into disjoint subsets, usually according to some
symmetry of the graph~\cite{Ge2011,AGT,BrouwerHaemers2012}. For unweighted
graphs (considered in this section), a partition of vertices 
$\Pi\eqdef\{C_i\}_{i=1}^m$ into $m$ disjoint vertex subsets or `cells' 
$C_i\subseteq V(\mathcal{G})$ (such that $V=\cup_iC_i$ and
$C_i\cap C_j=\varnothing$ for $i\neq j$) is said to be equitable if for 
every subset $C_i$ the number of edges $d_{ij}$ connecting $C_i$ to $C_j$ is 
independent of the choice 
of vertex in $C_i$. Thus, every vertex within a cell has the same number of 
neighbors in an adjacent cell, and equitable partitioning induces graph 
regularity within each cell and semi-regularity between cells. A cell 
containing only one vertex (i.e.\ $|C_i|=1$) is known as a `singleton cell.' 
Formal definitions and further details on equitable partitioning can be found 
in Appendix~\ref{equitableproof}.

Associating every cell to a single vertex yields a new graph which can be
considered as an exact coarse-grained version of the original graph (the 
interpretation of `exact' is given below); the resulting graph on $m$ vertices 
is known as the `quotient graph' $\Gamma\eqdef\mathcal{G}/\Pi$ 
under the equitable partitioning $\Pi$. The quotient graph derives its name 
from the elimination of eigenvalues from the spectrum, corresponding to 
dividing the characteristic equation of the graph adjacency matrix by a 
polynomial in $\tilde{\lambda}$ with the deleted eigenvalues as roots. For 
example, if the adjacency matrix $A_{\mathcal G}$ for a graph ${\mathcal G}$ 
with $n$ vertices has the characteristic polynomial 
$p_{\mathcal{G}}(\tilde{\lambda}) = (\tilde{\lambda}_1 - \tilde{\lambda})
(\tilde{\lambda}_2 - \tilde{\lambda})\cdots
(\tilde{\lambda}_N - \tilde{\lambda})$,
then at least one term is deleted in the adjacency matrix of the quotient 
graph, $p_{\Gamma}(\tilde{\lambda})
= (\tilde{\lambda}_1 - \tilde{\lambda})(\tilde{\lambda}_2 - \tilde{\lambda})
\cdots(\tilde{\lambda}_N - \tilde{\lambda})/(\tilde{\lambda}_2-\tilde{\lambda})
= (\tilde{\lambda}_1 - \tilde{\lambda})(\tilde{\lambda}_3 - \tilde{\lambda}) 
\cdots (\tilde{\lambda}_N - \tilde{\lambda})$. In practice, the partition 
process may eliminate more than one eigenvalue. It follows directly that the 
spectrum of the quotient graph is a subset of the spectrum of the primary 
graph~\cite{AGT}. Which eigenvalues remain under equitable partitioning depends
on the details, but in all cases the maximal eigenvalue is preserved. 
Equivalently, all the eigenvalues of the quotient (including the maximal one,
which corresponds to the ground-state of the physical system) are also 
eigenvalues of the original graph; the coarse-graining procedure is therefore
exact for (arguably) the most physically-relevant eigenvector. Further details 
can be found in Lemmas~\ref{lem:eqptEvalSubset} and 
\ref{lem:eqptAVecsToB} in Appendix~\ref{sec:EquitablePartitioning}. The edges 
between vertices of the quotient graph are generally weighted; the edge weights 
are given by (recall only for unweighted graphs)
\begin{equation}
\omega_{ij} = \sqrt{d_{ij}d_{ji}},\label{edgeweight}
\end{equation}
where $d_{ij}$ is the number of edges connecting cell $C_i$ to cell $C_j$. The 
weightings account for the semi-regularity of the connectivity between 
partitions. 

Equitable partitioning can be considered as an isometry effected by an operator
$Q$, known as the normalized partition matrix, defined as
\begin{equation}
Q = \sum_{i=1}^m\frac{1}{\sqrt{|C_i|}}\sum_{v\in C_i}|v\rangle
\langle\tilde{v}_i|,
\label{eq:unweightedQ}
\end{equation}
where $v\in V(\mathcal{G})$ and $\tilde{v}_i\in V(\Gamma)$. 
If $|V({\mathcal G})|=n$ and $|V(\Gamma)|=m$ then $Q$ must be a $n\times m$
matrix. The normalized partition matrix satisfies
the following properties~\cite{Godsil-periodic}:
\begin{eqnarray}
Q^TQ &=& I_m \nonumber\\ \left[
Q^TQ,A_{\mathcal{G}} \right] &=& \left[
QQ^T,A_{\mathcal{G}} \right] = 0.
\label{eq:q-prop1}
\end{eqnarray}
In general, $QQ^T\neq I_n$ unless every cell of $\Gamma$ is 
a singleton, so that $m=n$ and $Q=I_n$. While the first commutation
relation in Eq.~(\ref{eq:q-prop1}) is trivial, the second indicates that 
$QQ^T$ represents a symmetry of the graph $\mathcal{G}$.
The vertices are therefore grouped into cells according to this symmetry. 

The adjacency matrix of the quotient graph is obtained via
\begin{equation}
A_{\Gamma}\eqdef Q^TA_{\mathcal{G}}Q,
\label{eq:partition}
\end{equation}
i.e.\ the quotient graph is $\mathcal{G}$ with the attendant symmetry removed.
The spectral decomposition
\begin{equation}
A_{\Gamma}=\sum_i\tilde{\lambda}_iQ^T|z_i\rangle\langle z_i|
Q=\sum_i\tilde{\lambda}_i|\tilde{z}_i\rangle\langle\tilde{z_i}|
\label{eq:Aquotient}
\end{equation}
shows that the eigenvectors of the quotient graph adjacency matrix are 
related to those of the original graph via 
\begin{equation}
|\tilde{z}_i\rangle\eqdef Q^T|z_i\rangle.
\label{eq:vunderQ}
\end{equation}
This also reveals that the spectrum of the quotient is a subset of that of the
original graph, as expected from the definition of an equitable partition. 
Given that there are $n$ $|z_i\rangle$ eigenvectors but only $m$ 
$|\tilde{z}_i\rangle$ eigenvectors, some comments on Eqs.~(\ref{eq:Aquotient})
and (\ref{eq:vunderQ}) are in order. Because $QQ^T$ is a symmetry of the
adjacency matrix according to Eq.~(\ref{eq:q-prop1}), eigenvectors 
$|z_i\rangle$ are either even or odd eigenstates of $QQ^T$. Even functions map 
under $Q^T$ according to Eq.~(\ref{eq:vunderQ}), while odd functions map under 
$Q^T$ to the null vector. Thus, in practice, the sum in Eq.~(\ref{eq:Aquotient})
contains $m<n$ terms.

\begin{figure}
\begin{tikzpicture}

\filldraw[blue!50!white]		(0.75,0.75) 	circle (0.2)
								(0.75,-0.75)	circle (0.2)
								(1.5,0.75)		circle (0.2)
								(1.5,-0.75)		circle (0.2);
					
\filldraw[blue!50!white]		(0,0)			circle (0.2)
								(2.25,0)		circle (0.2);
					
\filldraw[blue!50!white]	 	(0.55,0.75) rectangle 	(0.95,-0.75)
								(1.3,0.75) 	rectangle	(1.7,-0.75);

\filldraw[black]	(0,0) 			circle (3pt)
					(0.75,0.75)		circle (3pt)
					(0.75,-0.75) 	circle (3pt)
					(0.75,0)		circle (3pt)
					(1.5,0.75) 		circle (3pt)
					(1.5,-0.75) 	circle (3pt)
					(1.5,0)	 		circle (3pt)
					(2.25,0)		circle (3pt)
					(5,0) 			circle (3pt)
					(6,0) 			circle (3pt)
					(7,0) 			circle (3pt)
					(8,0)			circle (3pt);

\draw (0,0)--(0.75,0.75);
\draw (0,0)--(0.75,-0.75);
\draw (0,0)--(0.75,0);
\draw (0.75,0.75)--(1.5,0.75);
\draw (0.75,0.75)--(1.5,0);
\draw (0.75,0)--(1.5,0.75);
\draw (0.75,0)--(1.5,-0.75);
\draw (0.75,-0.75)--(1.5,0);
\draw (0.75,-0.75)--(1.5,-0.75);
\draw (1.5,-0.75)--(2.25,0);
\draw (1.5,0)--(2.25,0);
\draw (1.5,0.75)--(2.25,0);

\draw (5,0)--(6,0);
\draw (6,0)--(7,0);
\draw (7,0)--(8,0);

\draw (5.5,0.5) node{$\sqrt{3}$};
\draw (6.5,0.5) node{$\sqrt{4}$};
\draw (7.5,0.5) node{$\sqrt{3}$};

\draw[->,very thick] (2.75,0).. controls (3.25,0.5) and (3.75,0.5) ..(4.5,0);
\draw (3.625,0.6) node{$Q$};

\end{tikzpicture}
\caption{(color online) Equitable mapping of the $3$-cube $\mathcal{Q}_3$ to 
the weighted line $\tilde{P}_4$. The hypercube $\mathcal{Q}_n$ on $n$ vertices 
has the weighted path graph as a quotient graph, with weights 
$w(v,v+1)=\sqrt{v(n+1-v)}$ symmetric about the center. The blue shaded regions
denote the cells in the equitable partitioning.}
\label{cube-mapping}
\end{figure}
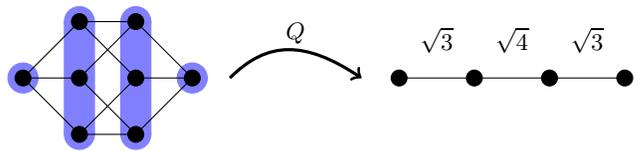

An example of equitable partitioning is shown in Fig.~\ref{cube-mapping}. One
can group vertices of the 3-cube into cells based on their distance from the 
left-most 
vertex. This yields a partitioning scheme with four cells, of which the two at
the ends are singletons. The normalized partition matrix can be written to 
clearly reveal the cell structure:
\begin{equation}
Q_{\mathcal{Q},4}=\left(\begin{matrix}
J_1 & {\zerovec}_1 & {\zerovec}_1 & {\zerovec}_1\cr
{\zerovec}_3 & J_3/\sqrt{3} & {\zerovec}_3 & {\zerovec}_3\cr
{\zerovec}_3 & {\zerovec}_3 & J_3/\sqrt{3} & {\zerovec}_3\cr
{\zerovec}_1 & {\zerovec}_1 & {\zerovec}_1 & J_1\cr
\end{matrix}
\right),
\end{equation}
where $J_k$ (${\zerovec}_k$) is the all-one (all-zero) vector of length $k$.
Applying this to the adjacency matrix for the hypercube via 
Eq.~(\ref{eq:partition}) 
yields the weighted path graph $\tilde{P}_4$ as a quotient, with successive 
weights $\sqrt{3}$, $\sqrt{4}$, and $\sqrt{3}$. In general, the equitable 
partitioning of $\mathcal{Q}_n$ to $\tilde{P}_{n+1}$ via the normalized
partition matrix $Q_{\mathcal{Q},n+1}$ yields the hypercubic edge weights 
$w(v,v+1)=\sqrt{v(n+1-v)}$. Just as the hypercube exhibits PST between 
antipodal vertices, so does the weighted path graph; this is an immediate 
consequence of the fact that antipodal vertices are singleton cells of the 
equitable partition~\cite{Tamon-Q}.

The standard definition of an equitable partition for an unweighted graph via 
the normalized partition matrix~(\ref{eq:unweightedQ}) must be extended to the 
case of weighted graphs in order to prove the main results of the present work.
This extension has not previously been made, to the best knowledge of the 
authors, and so it is outlined here with full definitions to be found in 
Appendix~\ref{equitableproof}. Define the weight of a vertex $v$ to be
\begin{equation}
\omega(v)\eqdef\sqrt{\sum_{u=1}^n A_{uv}^2},
\end{equation}
which in the unweighted case reduces to $\omega(v)=\sqrt{d_v}$, the square 
root of the degree of vertex $v$. Defining the normalized weight of $v$
with respect to its containing cell $C_{\tilde{v}}$ to be
\begin{equation}
\Omega(v) = \frac{\omega(v)}{\sqrt{\sum_{u\in C_{\tilde{v}}}\omega^2(u)}}
\eqdef\frac{\omega(v)}{\omega(C_{\tilde{v}})},
\label{eq:omega}
\end{equation} 
the normalized partition matrix for a weighted graph is then
\begin{equation}
Q=\sum_{v=1}^n\Omega(v)|v\rangle\langle\tilde{v}|.
\label{eq:Qweighted}
\end{equation}
Together with the constraints discussed in detail in
Appendix~\ref{equitableproof} that ensure a partition of a weighted graph
is equitable, this definition of the normalized partition matrix
is entirely consistent with the previous definition in the unweighted case.

Consider the time evolution operator for a quantum walker on the quotient 
graph,
\begin{equation}\label{eq:eqptSimplifyUB}
	U_{\Gamma}(t)
	\eqdef
	e^{-i Q^T A Q t}
	=
	\sum_{k=0}^\infty \frac{(-i t)^k}{k!}
		\left(Q^T A Q\right)^k.
\end{equation}
The $k=2$ term is
\begin{equation}
	\left(Q^TAQ\right)^2
	=
	Q^TAQQ^TAQ
	=
	Q^TA^2Q,
\end{equation}
where the last step follows from $[A,QQ^T]=0$ and $Q^TQ=I$. Because all powers
proceed in the same manner one obtains
\begin{equation}
	U_{\Gamma}(t) = Q^T U_{\mathcal{G}}(t) Q.
\label{eq:Uquotient}
\end{equation}

Result~(\ref{eq:Uquotient}) is particularly useful in the case of singleton 
cells. Suppose vertices $u$ and $v$ of $G$ are singletons in the partition
$\Pi$, in which case the maps from these vertices to their containing cells 
are invertible: $Q^T\ket{u}=\ket{\tilde{u}}$ and $Q\ket{\tilde{u}}=\ket{u}$.
The evolution between these vertices on the quotient graph is therefore 
identical to that between the associated vertices on the original graph:
\begin{equation}\label{eq:eqptCollapsedEvolution}
	\bra{\tilde{u}}U_{\Gamma}(t)\ket{\tilde{v}}
	=
	\bra{\tilde{u}}Q^T U_{\mathcal{G}}(t) Q\ket{\tilde{v}}
	=
	\bra{u}U_{\mathcal{G}}(t)\ket{v}.
\end{equation}
Equation~\eqref{eq:eqptCollapsedEvolution} holds for any undirected graph with 
real edge and self-loop weights under a partition that satisfies the condition 
for being equitable in the weighted case, 
Definition~\ref{def:eqptWeightedEquitable} in Appendix~\ref{equitableproof}.

\subsection{Multiboson walks on path graphs}
The graph-theoretic formalism for the analysis of many-boson systems has been
previously elucidated. The dynamics of quantum particles on discrete lattices
is governed by the hopping Hamiltonian, which is the negative of the graph
adjacency matrix (times an unimportant constant with units of energy). 
Non-interacting distinguishable bosons are represented by Cartesian powers of 
graphs, with identical bosons following by a suitable equitable 
partitioning~\cite{PST3}. Hard-core bosons are represented by symmetric powers 
of graphs, corresponding to deleting diagonal vertices of Cartesian powers, 
followed by an equitable partitioning~\cite{Audenaert2007}. Because the main 
result of the present work extends the concept of the symmetric power to 
weighted graphs, these two constructions are briefly reviewed here.

\subsubsection{Non-interacting bosons on weighted path graphs}
\label{sec:PSTnon}

The eigenvectors and eigenvalues of the hypercubically weighted path graphs 
(referred in this work generically as weighted path graphs) derive from those 
of the hypercubes. The eigenvectors are Krawtchouk polynomials 
$|\mathcal{K}_i\rangle$, which can be expressed in terms of hypergeometric 
functions or Jacobi polynomials~\cite{Feder-Krawchouk}. 
The eigenvalues of $A_{\mathcal{Q}_n}$ are the $n$-fold combinations of the
$\{-1,+1\}$ eigenvalues of $A_{P_2}$; thus 
$\tilde{\lambda}(A_{\mathcal{Q}_n})=\{-n,-n+2,\ldots,n-2,n\}$. 
All of these eigenvalues, except the extremal ones, are degenerate. Each of the
vertices in $\mathcal{Q}_n$ can be labeled by a unique bit string of length 
$n$. One can then construct a partition where cells are comprised of vertices
whose labels share the same Hamming weight. The resulting quotient graph has
the adjacency matrix $A_{\tilde{P}_{n+1}}$ of the weighted path graph. Its
spectrum coincides with that of $A_{\mathcal{Q}_n}$, i.e.\ 
$\tilde{\lambda}(A_{\tilde{P}_{n+1}})=-n+2\{0,1,\ldots,n\}$, but the degeneracy is completely 
removed. These eigenvalues obviously satisfy the ratio 
condition~(\ref{eq:ratio}). Alternatively, if one substitutes this spectrum 
into Eq.~(\ref{eq:UG2}) using $\tilde{\lambda}_i=\tilde{\lambda}_0=-n$, one 
obtains $U_{\mathcal{G}}(t)=e^{itn}\sum_je^{-2ijt} |z_j\rangle\langle z_j|$, 
which is equivalent to an identity (up to an unimportant constant) whenever
$t$ is an integer multiple of $\pi$. Thus, the weighted path graphs are 
periodic with period $\tau'=\pi$, a property inherited from the parent 
hypercubes.

Path graphs (both the hypercubically weighted $\tilde{P}_n$ and the simple 
$P_n$) on $n$ vertices are symmetric about their midpoints by construction. 
This ensures that all eigenvectors $|z_i\rangle$ of the associated adjacency 
matrices are either even or odd under a reflection about the midpoint of the 
graph, i.e.\ $|z_j\rangle\eqdef(z_{j,1}~z_{j,2}~\ldots~z_{j,n-1}~z_{j,n})^T
=\pm(z_{j,n}~z_{j,n-1}~\ldots~z_{j,2}~z_{j,1})^T
\eqdef\pm|\overline{z}_j\rangle$ where $z_{j,m}$ is the $m$th element of the 
$|z_j\rangle$ eigenvector and $|\overline{z}_j\rangle$ is $|z_j\rangle$ 
reflected about the midpoint. The upper (lower) sign corresponds to
an even (odd) eigenstate. Alternatively, the eigenvectors are eigenstates of 
the graph parity (or reflection) operator $\hat{P}$, where 
$\hat{P}|z_j\rangle=\pm|z_j\rangle$ for even (upper sign) and odd (lower sign) 
functions. Thus, $[\hat{P},A_{\tilde{P}_n}]=0$. Furthermore, for path graphs the
eigenvectors always satisfy $\hat{P}|z_j\rangle=(-1)^j|z_j\rangle$, so that the
parity of the eigenstates oscillates between even and odd as $j$ increases.

That PST occurs on hypercubically weighted path graphs follows directly from 
considerations of parity. PST for one particle on the weighted path graph 
corresponds to $U_{\mathcal{G}}(\tau)|z_j\rangle=e^{i\phi}|\overline{z}_j\rangle
=\pm e^{i\phi}|z_j\rangle=e^{i\phi}\hat{P}|z_j\rangle$, where $\phi$ is an 
unimportant global phase. Suppose that one constructs an equitable partitioning 
$\Pi_{\rm P}$ where each vertex is grouped in a cell with its reflection 
symmetry-related counterpart, i.e.\ vertex 1 with $n$, 2 with $n-1$, etc. 
Because the associated normalized partition matrix $Q_{\rm P}$ is even, 
$Q^T_{\rm P}|z_j\rangle=|\zerovec\rangle$ for $j$ odd. Every second eigenvalue
is thereby eliminated from the spectrum $\{-n,-n+4,\ldots,n-4,n\}$ of the 
quotient graph $\Gamma_P=\tilde{P}_n/\Pi_{\rm P}$. The quotient graph 
$\Gamma_P$ exhibits periodicity in time $\tau=\pi/2$. But every vertex state 
$|u\rangle$ in $\Gamma_P$ is a normalized superposition of symmetry-related 
vertices in $\tilde{P}_n$, i.e.\ $|u\rangle=\frac{1}{\sqrt{2}}\left(|i\rangle
+|n-i+1\rangle\right)$. Mapping $|u\rangle$ to itself in the quotient is 
equivalent to mapping $|i\rangle\leftrightarrow |n-i+1\rangle$, i.e.\ 
interchanging vertices across the midpoint in the original path graph. Thus, 
vertex periodicity at time $\tau'=\pi/2$ in the quotient implies PST in the 
original path graph at time $\tau=\pi/2$.

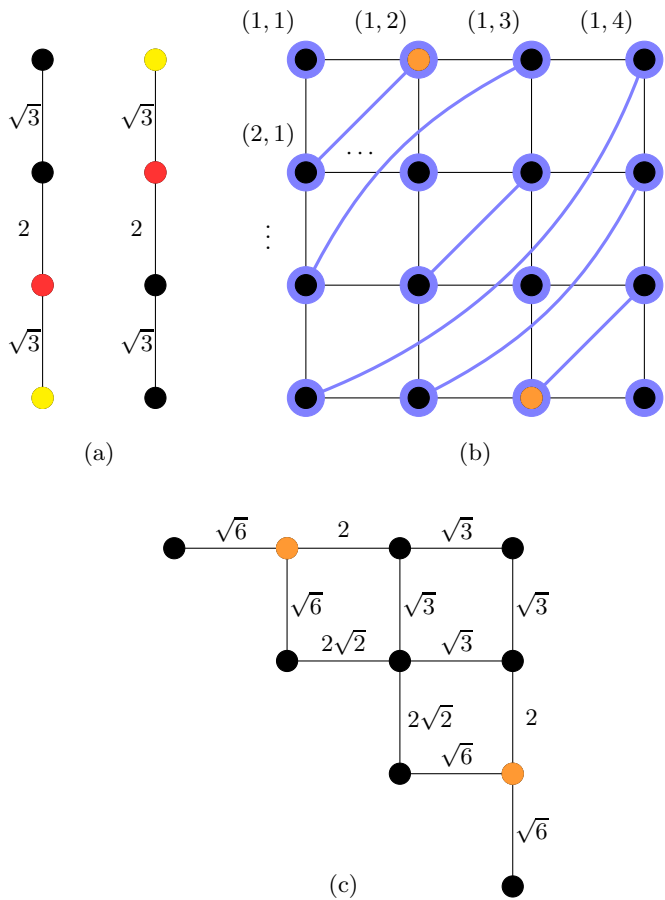
\begin{figure}
\begin{tikzpicture}
	\draw 	(3.5,0)--(8,0);
	\draw	(3.5,1.5)--(8,1.5);
	\draw	(3.5,3)--(8,3);
	\draw	(3.5,4.5)--(8,4.5);
	\draw 	(3.5,0)--(3.5,4.5);
	\draw	(5,0)--(5,4.5);
	\draw	(6.5,0)--(6.5,4.5);
	\draw	(8,0)--(8,4.5);
	\draw	(0,0)--(0,4.5);
	\draw	(1.5,0)--(1.5,4.5);
	\draw	(1.75,-2)--(6.25,-2);
	\draw	(3.25,-3.5)--(6.25,-3.5);
	\draw	(4.75,-5)--(6.25,-5);
	\draw	(6.25,-2)--(6.25,-6.5);
	\draw	(4.75,-2)--(4.75,-5);
	\draw	(3.25,-2)--(3.25,-3.5);

	\draw[blue!50!white,very thick](3.5,3)--(5,4.5);
	\draw[blue!50!white,very thick](5,1.5)--(6.5,3);
	\draw[blue!50!white,very thick](6.5,0)--(8,1.5);
	\draw[blue!50!white,very thick] (3.5,1.5).. controls (4.25,3) and (5,3.75) .. (6.5,4.5);
	\draw[blue!50!white,very thick] (3.5,0).. controls (5.75,0.85) and (7.25,2.35) .. (8,4.5);
	\draw[blue!50!white,very thick] (5,0).. controls (6.5,0.75) and (7.25,1.5) .. (8,3);
	\filldraw[blue!50!white] (3.5,0) circle (7pt);
	\filldraw[blue!50!white] (3.5,1.5) circle (7pt);
	\filldraw[blue!50!white] (3.5,3) circle (7pt);
	\filldraw[blue!50!white] (3.5,4.5) circle (7pt);
	\filldraw[blue!50!white] (5,0) circle (7pt);
	\filldraw[blue!50!white] (5,1.5) circle (7pt);
	\filldraw[blue!50!white] (5,3) circle (7pt);
	\filldraw[blue!50!white] (5,4.5) circle (7pt);
	\filldraw[blue!50!white] (6.5,0) circle (7pt);
	\filldraw[blue!50!white] (6.5,1.5) circle (7pt);
	\filldraw[blue!50!white] (6.5,3) circle (7pt);
	\filldraw[blue!50!white] (6.5,4.5) circle (7pt);
	\filldraw[blue!50!white] (8,0) circle (7pt);
	\filldraw[blue!50!white] (8,1.5) circle (7pt);
	\filldraw[blue!50!white] (8,3) circle (7pt);
	\filldraw[blue!50!white] (8,4.5) circle (7pt);

	\filldraw[black]	(3.5,0) circle (4pt)
						(3.5,1.5) circle (4pt)
						(3.5,3) circle (4pt)
						(3.5,4.5) circle (4pt)
						(5,0) circle (4pt)
						(5,1.5) circle (4pt)
						(5,3) circle (4pt)
						(5,4.5) circle (4pt)
						(6.5,0) circle (4pt)
						(6.5,1.5) circle (4pt)
						(6.5,3) circle (4pt)
						(6.5,4.5) circle (4pt)
						(8,0) circle (4pt)
						(8,1.5) circle (4pt)
						(8,3) circle (4pt)
						(8,4.5) circle (4pt)
						(0,0) 		circle (4pt)
						(0,1.5) 	circle (4pt)
						(0,3) 		circle (4pt)
						(0,4.5) 	circle (4pt)
						(1.5,0)		circle (4pt)
						(1.5,1.5)	circle (4pt)
						(1.5,3) 	circle (4pt)
						(1.5,4.5)	circle (4pt)
						
						(1.75,-2) circle (4pt)
						(3.25,-2) circle (4pt)
						(4.75,-2) circle (4pt)
						(6.25,-2) circle (4pt)
						(3.25,-3.5) circle (4pt)
						(4.75,-3.5) circle (4pt)
						(6.25,-3.5) circle (4pt)
						(4.75,-5) circle (4pt)
						(6.25,-5) circle (4pt)
						(6.25,-6.5) circle (4pt);

	\filldraw[yellow] (0,0) circle (4pt);
	\filldraw[red!80!white] (0,1.5) circle (4pt);
	
	\filldraw[red!80!white] (1.5,3) circle (4pt);
	\filldraw[yellow] (1.5,4.5) circle (4pt);
	
	\filldraw[orange!80!white] (5,4.5) circle (4pt);
	\filldraw[orange!80!white] (6.5,0) circle (4pt);
	
	\filldraw[orange!80!white] (3.25,-2) circle (4pt);
	\filldraw[orange!80!white] (6.25,-5) circle (4pt);

		\draw (2.5,-1.75)	node{$\sqrt{6}$};
		\draw (4,-1.75)	node{$2$};
		\draw (5.5,-1.75)	node{$\sqrt{3}$};

		\draw (4,-3.25)	node{$2\sqrt{2}$};
		\draw (5.5,-3.25)	node{$\sqrt{3}$};

		\draw (5.5,-4.75)	node{$\sqrt{6}$};

		\draw (3.5,-2.75)	node{$\sqrt{6}$};		
		\draw (5,-2.75)	node{$\sqrt{3}$};
		\draw (6.5,-2.75)	node{$\sqrt{3}$};

		\draw (5.15,-4.25)	node{$2\sqrt{2}$};
		\draw (6.5,-4.25)	 node{$2$};
		
		\draw (6.5,-5.75)	node{$\sqrt{6}$};
	
		\draw (3,5) node{$(1,1)$};
		\draw (4.5,5)	 node{$(1,2)$};
		\draw (6,5)  node{$(1,3)$};
		\draw (7.5,5)  node{$(1,4)$};
		\draw (3,3.5) node{$(2,1)$};
		\draw (4.25,3.25) node{$\cdots$};
		\draw (3,2.25) node{$\vdots$};
		
		\draw	(-0.25,0.75) node{$\sqrt{3}$};
		\draw	(-0.25,2.25)   node{$2$};
		\draw	(-0.25,3.75) node{$\sqrt{3}$};
		
		\draw	(1.25,0.75) node{$\sqrt{3}$};
		\draw	(1.25,2.25)   node{$2$};
		\draw	(1.25,3.75) node{$\sqrt{3}$};
		
		\draw (0.75,-0.75) node{(a)};
		\draw (5.75,-0.75) node{(b)};
		\draw (4,-6.5) node{(c)};

\end{tikzpicture}
\caption{(color online) Two copies of the hypercubically weighted path graph on
four vertices $\tilde{P}_4$ are shown in (a). Suppose that two distinguishable
particles (yellow and red) are incident on the first and second sites; after
undergoing unitary evolution for a time $\tau=\pi/2$, they are located at the
fourth and third sites, respectively. The graph representation of their joint
occupation space configuration is shown in (b), with indices labeling the 
possible positions of the two particles. The evolution described in (a) is 
represented by two orange-colored vertices of a single boson; the initial 
state is $(1,2)$, and the final state is $(4,3)$. The cell structure of 
the equitable partitioning from distinguishable to identical particles
is depicted via the blue lines connecting vertices which are mirror symmetric 
with respect to the central diagonal (i.e.\ equivalent under a permutation of
the indices). The resulting quotient graph representing two identical bosonic 
particles is shown in (c). The edge weights reflect the action of generalized 
equitable partitioning.}
\label{config-NI}
\end{figure}

It has been shown in Ref.~\cite{PST3} that the evolution of $k$
distinguishable non-interacting particles on a graph $\mathcal{G}$ is 
physically equivalent to the evolution of a single walker on the $k$-fold 
Cartesian power of $\mathcal{G}$, denoted $\mathcal{G}^{\Box k}$. 
Figure~\ref{config-NI} shows an example of two particles on the path graph
$\tilde{P}_4$ with hypercubic edge weights. Figure~\ref{config-NI}(a) shows 
two copies of $\tilde{P}_4$, each with a configuration two particles. The 
two-particle effective graph $\tilde{P}_4^{\Box 2}$ is shown in 
Fig.~\ref{config-NI}(b). If $|z_i\rangle$ is an eigenvector of 
$A_{\mathcal{G}}$ with eigenvalue $\tilde{\lambda}_i$, then using 
Eq.~(\ref{eq:CartesianProduct}) one obtains
\begin{eqnarray}
A_{\mathcal{G}^{\Box 2}}|z_i\rangle\otimes|z_j\rangle &=& \left( 
A_{\mathcal{G}}\otimes I_n + I_n\otimes A_{\mathcal{G}} \right)\left( 
|z_i\rangle\otimes|z_j\rangle \right) \nonumber\\ &=& 
\left( A_{\mathcal{G}}|z_i\rangle\otimes 
|z_j\rangle + |z_i\rangle\otimes A_{\mathcal{G}}|z_j\rangle \right) 
\nonumber\\ &=& \left( 
\tilde{\lambda}_i|z_i\rangle\otimes|z_j\rangle + 
|z_i\rangle\otimes\tilde{\lambda}_j|z_j\rangle \right) 
\nonumber\\ &=& \left(\tilde{\lambda}_i + \tilde{\lambda}_j\right)\left( 
|z_i\rangle\otimes|z_j\rangle \right),
\end{eqnarray}
revealing that the eigenvalues of $k$-fold Cartesian powers of $\mathcal{G}$
correspond to all possible additive combinations of $k$ eigenvalues. Recall
from Sec.~\ref{sec:backwalk} that if the graph $\mathcal{G}$ exhibits PST then
so does its Cartesian power, and the eigenvalues of $A_{\mathcal{G}}$ 
automatically satisfy the ratio condition~(\ref{eq:ratio}). Likewise, the 
eigenstates of Cartesian powers of path graphs remain eigenstates of the global 
parity operator.

When the particles are instead considered to be indistinguishable bosons, any 
vertices in the Cartesian power with labels that are identical under a 
permutation are grouped together within a single cell. The associated
equitable partitioning is effected by the normalized partition matrix 
$Q_{\rm I}$, where the `I' subscript denotes `indistinguishability.' Each of 
the $n^k$ vertices $v$ of the $k$th Cartesian power of $\mathcal{G}$ is labeled 
by a list of length $k$, given as $v=(x_1, x_2,\cdots, x_k)$, where 
$x_i\in\{ 1, 2, \cdots, n \}$. The partition $\Pi_{\rm I}$ groups together all 
vertices that are permutations $p(v)$ of the list $v$ into a single cell. 
Figure~\ref{config-NI}(b)
shows the equitable partitioning for two bosons on $\tilde{P}_4$. For example,
vertices with the labels $(1,2)$ and $(2,1)$ constitute a unique cell. The
quotient graph is shown in Fig.~\ref{config-NI}(c). Its eigenvalues are a 
subset of all combinations of those of $A_{\tilde{P}_4}$ by 
Lemmas~\ref{lem:eqptEvalSubset} and \ref{lem:eqptAVecsToB} in 
Appendix~\ref{sec:EquitablePartitioning}, and therefore still satisfy the ratio 
condition~(\ref{eq:ratio}). The vertices derived from 
singleton cells exhibit PST via Eq.~(\ref{eq:eqptCollapsedEvolution}). All 
other vertices are derived from cells with exactly two vertices; each of these 
two vertices in the Cartesian product graph exhibits PST to one of two vertices 
both located in a distinct cell. Thus, all vertices in the quotient graph also 
exhibit PST between any two vertices with labels
$(i,j,\ldots,k)$ and $(n+1-i,n+1-j,\ldots,n+1-k)$ in time $\tau=\pi/2$. 

\subsubsection{Hard-core bosons on path graphs}
\label{sec:interactions}

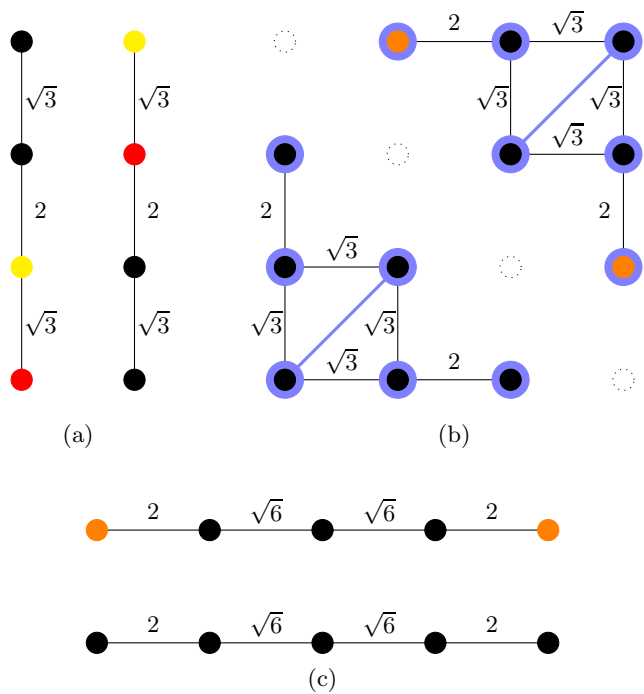
\begin{figure}
\begin{tikzpicture}

		\draw (3.5,0)--(3.5,3);
		\draw (5,0)--(5,1.5);
		\draw (3.5,0)--(6.5,0);
		\draw (3.5,1.5)--(5,1.5);
		
		\draw (5,4.5)--(8,4.5);
		\draw (6.5,3)--(8,3);
		\draw (6.5,3)--(6.5,4.5);
		\draw (8,1.5)--(8,4.5);
		
		\draw (0,0)--(0,4.5);
		\draw (1.5,0)--(1.5,4.5);
		
		\draw (1,-2)--(7,-2);
		\draw (1,-3.5)--(7,-3.5);
		
		\draw[blue!50!white, very thick] (8,4.5)--(6.5,3);
		\draw[blue!50!white, very thick] (5,1.5)--(3.5,0);
		\filldraw[blue!50!white]	(8,4.5) 	circle (7pt)
									(6.5,3) 	circle (7pt)
									(6.5,4.5) 	circle (7pt)
									(8,3)		circle (7pt)
									(8,1.5) 	circle (7pt)
									(5,4.5) 	circle (7pt)
									(5,1.5) 	circle (7pt)
									(3.5,0) 	circle (7pt)
									(5,0)   	circle (7pt)
									(6.5,0) 	circle (7pt)
									(3.5,1.5) 	circle (7pt)
									(3.5,3) 	circle (7pt);

	\filldraw[black]	(0,3) 		circle (4pt)
						(0,4.5) 	circle (4pt)
						
						(1.5,0) 	circle (4pt)
						(1.5,1.5) 	circle (4pt)
						
						(3.5,0) 	circle (4pt)
						(3.5,1.5) 	circle (4pt)
						(3.5,3) 	circle (4pt)
						(5,0) 		circle (4pt)
						(5,1.5)		circle (4pt)
						(6.5,0) 	circle (4pt)
						
						(6.5,4.5) 	circle (4pt)
						(6.5,3) 	circle (4pt)
						(8,4.5)		circle (4pt)
						(8,3) 		circle (4pt)
						
						(2.5,-2)	circle (4pt)
						(4,-2)		circle (4pt)
						(5.5,-2) 	circle (4pt)
						
						(1,-3.5)	circle (4pt)
						(2.5,-3.5)	circle (4pt)
						(4,-3.5)	circle (4pt)
						(5.5,-3.5) 	circle (4pt)
						(7,-3.5)	circle (4pt);
						
	\draw[black,dotted]	(3.5,4.5)	circle (4pt)
						(5,3)		circle (4pt)
						(6.5,1.5)	circle (4pt)
						(8,0)		circle (4pt);

\filldraw[orange]		(5,4.5)		circle (4pt)
						(8,1.5)		circle (4pt);
						
\filldraw[red]			(0,0)		circle (4pt)
						(1.5,3)		circle (4pt);
						
\filldraw[yellow]		(0,1.5)		circle (4pt)
						(1.5,4.5) 	circle (4pt);
						
\filldraw[orange]		(1,-2)		circle (4pt)
						(7,-2)		circle (4pt);

	\draw (0.25,0.75) 	node{$\sqrt{3}$};
	\draw (0.25,2.25) 	node{$2$};
	\draw (0.25,3.75)	node{$\sqrt{3}$};

	\draw (1.75,0.75) 	node{$\sqrt{3}$};
	\draw (1.75,2.25)	node{$2$};
	\draw (1.75,3.75)	node{$\sqrt{3}$};

	\draw (3.25,0.75) node{$\sqrt{3}$};
	\draw (3.25,2.25) node{$2$};
	\draw (4.75,0.75) node{$\sqrt{3}$};
	\draw (4.25,1.75) node{$\sqrt{3}$};
	\draw (4.25,0.25) node{$\sqrt{3}$};
	\draw (5.75,0.25) node{$2$};
	
	\draw (5.75,4.75) node{$2$};
	\draw (7.25,4.75) node{$\sqrt{3}$};
	\draw (7.25,3.25) node{$\sqrt{3}$};
	\draw (7.75,2.25) node{$2$};
	\draw (7.75,3.75) node{$\sqrt{3}$};
	\draw (6.25,3.75) node{$\sqrt{3}$};

	\draw (1.75,-1.75)	node{$2$};
	\draw (3.25,-1.75)	node{$\sqrt{6}$};
	\draw (4.75,-1.75)	node{$\sqrt{6}$};
	\draw (6.25,-1.75)	node{$2$};

	\draw (1.75,-3.25)	node{$2$};
	\draw (3.25,-3.25)	node{$\sqrt{6}$};
	\draw (4.75,-3.25)	node{$\sqrt{6}$};
	\draw (6.25,-3.25)	node{$2$};

	\draw (0.75,-0.75) node{(a)};
	\draw (5.75,-0.75) node{(b)};
	\draw (4,-4)	   node{(c)};

\end{tikzpicture}
\caption{(color online) The effective graph representing two hard-core
distinguishable bosons on the hypercubically weighted path graph $\tilde{P}_4$. Two 
copies of $\tilde{P}_4$ are shown in (a). The left shows two distinguishable
hard-core particles (yellow and red) incident on the first and second sites; 
after undergoing unitary evolution for a time $\tau=\pi/2$, they are located 
sites three and four, respectively (note that unlike noninteracting particles
they cannot exchange their relative positions). The graph representation of 
their joint occupation space configuration is shown in (b). This is the 
Cartesian square $\tilde{P}_4^{\Box 2}$ with vertices deleted whose indices 
have repeating labels, leading to two disconnected graphs (deleted vertices are
indicated by the dashed circles on the diagonal). Input and output vertices 
under PST, corresponding to locations in (a), are shown in orange. The quotient 
disconnected graph components for identical hard-core bosons are shown in (c).}
\label{hard-core-PST}
\end{figure}

Thus far, we have ignored all particle interactions. When the repulsion between 
two particles is so strong that there is an infinite energy cost to occupying
the same vertex, then the particles are said to have `hard-core' interactions.
In this case, at most one particle can occupy any given vertex at a time; this
restriction is strongly reminiscent of the Pauli exclusion principle for 
(spinless) fermions, a correspondence that will be made explicit in 
Sec.~\ref{sec:TG}. Hard-core particles undergoing a quantum walk cannot pass through each
other, which implies that their initial ordering on path graphs cannot change 
as a function of time. This behavior is illustrated in 
Fig.~\ref{hard-core-PST}(a).

In the graph theory
context, local (i.e.\ on-site) interactions are represented by self-loops on 
vertices whose labels have at least one repeating index
(or, in the indistinguishable case, those corresponding to multiple occupancy).
An example of such a
vertex label is the sequence $(1,2,2,3,\ldots)$. In the limit of infinitely
strong interactions (either positive or negative), the hard-core limit, these 
vertices are deleted from the graph. Physically, particles will never occupy a 
vertex with a local infinite positive strength; alternatively, any particle 
incident on a vertex with infinite negative strength would never leave.

The vertex deletion operator $D$ is a diagonal matrix with entries of either 
zero or one; the zeroes are at vertices to be deleted. The hard-core adjacency 
matrix is obtained directly from that of the non-interacting case via
\begin{equation}
A_{\rm HC}\eqdef DA_{\rm NI}D,
\label{eq:deletion}
\end{equation}
where $A_{\rm HC}$ and $A_{\rm NI}$ represent the 
adjacency matrices of the hard-core and the non-interacting graphs 
respectively. Figure~\ref{hard-core-PST}(b) depicts the effect of the deletion 
matrix on the $\tilde{P}_4^{\Box 2}$ graph shown in Fig.~\ref{config-NI}(b).
In this case, the vertex deletion yields two disconnected graph components because the
full diagonal of the original Cartesian product is deleted. Importantly for the
hard-core bosons on path graphs, the two graph components are in fact 
isomorphic. As will be shown in detail in this work, the ability of the deleted 
graph to support PST hinges on the fact that the deleted graph is composed of 
disconnected isomorphic graph components.

The adjacency matrix for $k$ identical hard-core bosons on a graph
$\mathcal{G}$ can therefore be defined as
\begin{equation}
A_{\rm HC,I}\eqdef Q_{\rm I}^TDA_{\mathcal{G}^{\Box k}}DQ_{\rm I},
\label{eq:SymmetricPower}
\end{equation}
where $Q_{\rm I}$ is the normalized partition matrix effecting the map from 
distinguishable to identical particles, as illustrated in the map 
from Fig.~\ref{config-NI}(b) to \ref{config-NI}(c) for bosons on $\tilde{P}_4$.
The definition~(\ref{eq:SymmetricPower}) is identical to the $k$th symmetric 
power $\mathcal{G}^{\{k\}}\equiv((\mathcal{G}^{\Box k})\backslash D)
/\Pi_{\rm I}$~\cite{Audenaert2007}. 
Because $D$ and $A_{\mathcal{G}^{\Box k}}$ 
do not commute, the eigenvalues of symmetric powers are not trivially related 
to those of the Cartesian powers from which they are derived. In fact, as 
discussed further in the following, even if $\mathcal{G}$ exhibits PST, 
$\mathcal{G}^{\{k\}}$ generally does not. 

Note that the deletion matrix need not be applied before equitably 
partitioning. An alternative approach is to first obtain the generalized 
quotient graph for identical noninteracting particles on the weighted
path graph, and then delete vertices corresponding to multiple occupancy. For 
example, deleting vertices along the bottom diagonal in the graph shown in 
Fig.~\ref{config-NI}(c) yields the same weighted graph components shown in 
Fig.~\ref{hard-core-PST}(b).

In Fig.~\ref{hard-core-PST}(b), each disconnected graph corresponds to a 
different ordering of the bosons on the underlying path graph. This has two 
important implications. First, applying the equitable partitioning for
indistinguishability yields exactly one of the disconnected graphs as the
resulting symmetric power graph. Second, if there is PST for hard-core bosons,
it must be between two vertices within a given disconnected graph. This is in
marked contrast with the situation for non-interacting bosons, where the PST
necessarily traverses the now-deleted vertices. Figure~\ref{hard-core-PST}(a) 
illustrates this for two hard-core bosons on $\tilde{P}_4$. Because the two
bosons cannot exchange their positions (doing so would require them to occupy 
the same site), the ordering of the bosons must remain the same.

Figure~\ref{hard-core-PST}(c) shows the results of equitably partitioning the
disconnected graphs depicted in Fig.~\ref{hard-core-PST}(b), where each cell
in the partition is either a singleton or given by a pair of blue circles connected by blue lines. The orange 
vertices in Figs.~\ref{hard-core-PST}(b) and \ref{hard-core-PST}(c) represent 
the (equivalent) sites between which PST would occur. It is interesting to note
that the quotient of the graph for two identical hard-core bosons on 
$\tilde{P}_4$, shown in Fig.~\ref{hard-core-PST}(c), is itself equivalent to 
the graph for a single particle on the hypercubically weighted path graph 
$\tilde{P}_5$ which exhibits PST by construction. This observation was an 
initial motivation for the present work, as this ensures PST for two hard-core 
bosons on $\tilde{P}_4$. Unfortunately, for two hard-core bosons on larger 
weighted path graphs $\tilde{P}_m$, $m>4$, there is no such simple equitable 
partition that can yield a $\tilde{P}_q$ quotient for some choice of $q$; 
likewise for larger numbers of particles. This raises the question of whether 
PST is generally possible for $k$ particles on $\tilde{P}_n$.

\subsection{Tonks-Girardeau solution}
\label{sec:TG}

An exact solution is known for a system of $k$ hard-core bosons in one 
dimension, known as the Tonks-Girardeau (TG) {\it ansatz}~\cite{TG,TG2}. The 
solution consists of a mapping from hard-core bosons to non-interacting 
spinless fermions. The deletion of vertices associated with multiple occupancy
is then automatically accounted for by the Pauli exclusion principle, and so
no explicit application of a deletion operator is required. The Pauli exclusion
principle requires that at most one particle can occupy a given vertex. The 
fermionic states are then explicitly symmetrized in order to be consistent with 
bosonic symmetry. This is accomplished through the use of a `unit antisymmetry 
operator' $\mathcal{A}$, a diagonal matrix with $\pm 1$ entries.
The validity of the fermionic mapping hinges on the requirement that
\begin{equation}
\left[ A_{P_n^{\Box k}} ,\mathcal{A} \right] = 0,
\label{eq:commutativity}
\end{equation}
which will be proved in Lemmas~\ref{lem:Ablocks} and \ref{lem:fermibose} below.
This guarantees that symmetric versions of the fermionic eigenstates remain
eigenstates of the Cartesian power, and furthermore that these states are 
mapped to eigenstates of the (bosonic) quotient graph via $Q^T$.

The TG construction proceeds as follows. There are $d=n!/k!(n-k)!$ arrangements 
of $k$ indistinguishable fermionic 
particles on a graph ${\mathcal G}$ with $n$ vertices. Because the fermions are 
assumed to be non-interacting, each of the $d$ eigenstates of the $k$-fermion
system consists of each particle occupying a distinct eigenstate of 
$A_{\mathcal G}$. One can label each of these $d$ states by an ordered 
$k$-tuple with elements from $\{0,\ldots,n-1\}$
labeling the (ordered) eigenstates of 
$A_{\mathcal G}$, and define $\ell$ the cardinality-$d$ set of all such 
tuples; for example, the set $\ell$ has elements 
$\ell_1=(0,1,\ldots,k-2,k-1)$ corresponding to particles in the 
first $k$ single-particle eigenstates of $A_{\mathcal G}$, 
$\ell_2=(0,1,\ldots,k-2,k)$, and $\ell_d=(n-k,n-k+1,\ldots,n-1)$. One can
assume that $k\leq n/2$ because the states for $k>n/2$ are equivalent to those
with $k<n/2$ by particle-hole symmetry. The fermionic states for $k$ particles 
on $P_n$ are therefore
\begin{equation}
|\mathcal{F}_i\rangle = \sum\limits_{p\in S_{k}}\sigma\left[p\left(\ell_i\right)
\right]\bigotimes_{j=1}^k|z_{\ell_i(j)}\rangle,
\label{eq:fermifunction}
\end{equation}
where $p$ represents a permutation of the elements of 
$\ell_j$, $S_{k}$ represents the symmetric group of the $k$ eigenvectors 
$|z_{\ell_i(j)}\rangle$ of $P_n$, and $\ell_i(j)$ denotes the $j$th element of
the ordered $k$-tuple $\ell_i$. The $\sigma$ function of the permutation $p$ 
returns the sign of $p$, and ensures the required antisymmetric behaviour of 
the fermionic state. The TG solution for the eigenvectors of hard-core bosons 
on (unweighted) path graphs is therefore
\begin{equation}
|{\mathcal B}_i\rangle={\mathcal A}|{\mathcal F}_i\rangle,
\label{eq:TG}
\end{equation}
where ${\mathcal A}$ is the unit antisymmetry operator ensuring that all
fermionic eigenvectors $|{\mathcal F}_i\rangle$, defined in 
Eq.~(\ref{eq:fermifunction}), are symmetric on particle interchange.

The eigenvalues $\lambda_i$ of the adjacency matrix for $k$ free fermions on a 
graph $\mathcal{G}$ coincide with those of the Cartesian power of 
$\mathcal{G}$, i.e.\ combinations of eigenvalues $\tilde{\lambda}_i$ of 
$A_{\mathcal{G}}$ (the antisymmetrization of the eigenstates does not affect
the eigenvalues). The ratio condition~(\ref{eq:ratio}) then takes the form
\begin{equation}
\frac{%
\sum_{i\in \Lambda_p}\tilde{\lambda}_i%
-\sum_{i\in \Lambda_q}\tilde{\lambda}_i%
}{%
\sum_{i\in \Lambda_r}\tilde{\lambda}_i%
-\sum_{i\in \Lambda_s}\tilde{\lambda}_i%
} \in \mathbb{Q},
\label{ratio-HC}
\end{equation}
where each $\Lambda$ is a subset of $k$ eigenvalues of $\mathcal{G}$.
Within the possible forms of Eq.~\eqref{ratio-HC} exists a subset for which
the sums in the numerator differ in only a single term, and likewise for the
denominator:
\begin{equation}
\frac{%
\sum_{i\in \Lambda_p}\tilde{\lambda}_i%
-\sum_{i\in \Lambda_q}\tilde{\lambda}_i%
}{%
\sum_{i\in \Lambda_r}\tilde{\lambda}_i%
-\sum_{i\in \Lambda_s}\tilde{\lambda}_i%
}%
= \frac{%
\tilde{\lambda}_n - \tilde{\lambda}_m%
}{%
\tilde{\lambda}_x - \tilde{\lambda}_y%
}\in\mathbb{Q},
\label{eq:fin-ratio}
\end{equation}
where the remaining $\tilde{\lambda}_{n,m,x,y}$ can each correspond to any 
eigenvalue of $A_{\mathcal{G}}$ (subject to
$\tilde{\lambda}_x\neq\tilde{\lambda}_y$).
Thus, if the ratio condition holds for 
$A_{\mathcal{G}}$
then it necessarily holds for $k$ fermions on $\mathcal{G}$, and vice versa. 
Likewise, if Eq.~(\ref{eq:commutativity}) is satisfied then so is the ratio condition
for hard-core bosons. Satisfying the ratio condition is only necessary 
for PST, however, not sufficient. That PST indeed occurs with hard-core bosons 
on weighted path graphs is the central result of the present work, and is 
proven in Sec.~\ref{sec:results}.

\section{Results}
\label{sec:results}

In this section we prove that there exists PST for $k$ distinguishable or 
identical hard-core bosons on the hypercubically weighted graphs
$\tilde{P}_n$. This is done in two ways, first through the direct application 
of the PST definition from Eq.~\eqref{eq:PST}, and second using the properties
of equitable partitioning in order to draw conclusions about the 
characteristics of the quotient graph. Before proceeding to the main proofs,
the graph structure of hard-core bosons is elucidated.

\subsection{Disconnected graphs under deletion}

An important property of the adjacency matrix for $k$ distinguishable hard-core 
bosons on $P_n$ or $\tilde{P}_n$ is that it disconnects into $k!$ separate 
graph components, all of which are isomorphic to one another. This is shown in
Lemma~\ref{lem:Ablocks}. 
\begin{lemma}
The graph of $k$ distinguishable hard-core bosons on a path graph of length $n$ ($P_n$
or $\tilde{P}_n$) consists of $k!$ isomorphic disconnected 
components described by identical adjacency matrices $A_{{\rm HC},i}$:
$DA_{P_n^{\Box k}}D=\oplus_{i=1}^{k!}A_{{\rm HC},i}$
\label{lem:Ablocks}
\end{lemma}

\begin{prooflem}

The graph for $k$ distinguishable non-interacting particles on a path graph is 
$P_n^{\Box k}$. Each of the $n^k$ vertices $v$ is labeled by a list of length 
$k$, given as $v=(x_1, x_2,\cdots, x_k)$, where
$x_i\in\{ 1, 2, \cdots, n \}$. Consider two vertices 
$v_1=(x_1, \cdots, x_i, x_j,\cdots, x_k)$ and 
$v_2=(x_1, \cdots, x_j, x_i,\cdots, x_k)$, where $x_i > x_j$. The 
shortest path from $v_1$ to $v_2$ corresponds to the sequence 
$(x_i,x_j)$, $(x_i-1,x_j)$, $\ldots$, $(x_j,x_j)$, $(x_j,x_j+1)$, $\ldots$, 
$(x_j, x_i)$, with the remaining vertex labels held constant.
If $D$ deletes all vertices whose labels have repeating indices, 
then the intermediate vertex $v=\{ x_1, \cdots, x_j, x_j,\cdots, x_k \}$ in 
this sequence is disconnected from both $v_1$ and $v_2$. 

The positions of the 
entries $x_i$ and $x_j$ must interchange on the path from $v_1$ to $v_2$, but 
any edge increments $x_i\to x_i\pm 1$ for only one entry of the list 
(subject to $1\leq x_i\pm1\leq n$). This ensures that for any longer path at least
one of the traversed vertices will have a pair of repeating indices, which is
deleted by $D$. Therefore $v_1$ and $v_2$ are disconnected. The map from 
$v_1$ to $v_2$ can equivalently be represented by an index permutation $p$. 
For an index list of length $k$, there are $k!$ possible permutations of 
the indices. Therefore, for $k$ hard-core bosons on a path graph of length $n$, 
there will be $k!$ induced graph components.

There are $n!/k!(n-k)!$ ordered lists of length $k$ with non-repeating 
elements drawn from the set $\{1,\ldots,n\}$. There are $k!$ 
permutations of each of these, yielding $n!/(n-k)!$ lists with non-repeating
elements. Because all other lists have at least one repeating element, $D$
deletes a total of $n^k-n!/(n-k)!$ vertices from the Cartesian power graph.
Thus, each of the $k!$ graph components is of size $d\eqdef n!/k!(n-k)!$.
It is therefore always possible to write
\begin{equation}
A_{\rm HC}=DA_{P_n^{\Box k}}D=\bigoplus_{i=1}^{k!}A_{{\rm HC},i},
\label{eq:block_adj}
\end{equation}
where $A_{{\rm HC},i}$ are the $d\times d$ adjacency matrices of the $k!$ graph 
components. 

Furthermore, each vertex $v$ in a given graph component has a unique vertex 
counterpart in each of the $k!-1$ other graph components, labeled by $p(v)$, 
where $p$ is a permutation of the list elements. The same $p$ maps all vertices
in a given component to all vertices in a different component, which makes it
an isomorphism. All $A_{{\rm HC},i}$ are therefore isomorphic, which implies
that there exists an operator $\hat{\pi}_{i,j}$ satisfying 
$\hat{\pi}_{i,j}\hat{\pi}_{i,j}^T=\hat{\pi}_{i,j}^T\hat{\pi}_{i,j}=I$, derived 
from $p\equiv p_{ij}$, such that 
$\hat{\pi}_{i,j}A_{{\rm HC},j}\hat{\pi}_{i,j}^T=A_{{\rm HC},i}$, $i\neq j$.

\end{prooflem}

With this lemma in hand, it is straighforward to prove that one can always
express the wavefunction for distinguishable hard-core bosons on path graphs as
the symmetrized wavefunction for non-interacting fermions.

\begin{lemma}
(Tonks and Girardeau~\cite{TG,TG2}) The $j$th wavefunction 
$|\mathcal{B}_j\rangle$ for $k$ distinguishable 
hard-core bosons on a path graph of length $n$ ($P_n$ or $\tilde{P}_n$) can
be expressed in terms of the wavefunction  $|\mathcal{F}_j\rangle$
for $k$ non-interacting fermions on the same graph via
\begin{equation}
|\mathcal{B}_j\rangle=\mathcal{A}|\mathcal{F}_j\rangle,
\label{eq:bosefermi}
\end{equation}
where the unit antisymmetry operator is defined as
\begin{equation}
\mathcal{A}\eqdef\bigoplus_{i=1}^{k!}\sigma(p_{i,1})I_d.
\label{eq:Ablockdiag}
\end{equation}
Here $I_d$ is the
identity matrix of size $d$ and $\sigma$ returns the sign of the vertex
permutation $p_{i,1}$ from the graph component labeled `1' to that labeled
`$i$'.
\label{lem:fermibose}
\end{lemma}

\begin{prooflem}

The direct-sum structure of the hard-core adjacency matrix~(\ref{eq:block_adj}) 
implies that the eigenvectors of $A_{\mathrm{HC}}$ follow directly from those
of $A_{{\rm HC},1}$, given by
\begin{equation}
A_{{\rm HC},1}|y_\lblvec^{(1)}\rangle
= \lambda_\lblvec^{(1)}|y_\lblvec^{(1)}\rangle,
\label{eq:eig-def}
\end{equation}
where $\lblvec\in\{1,\ldots,d\}$.
There are $k!d$ eigenvectors of $A_{\rm HC}$;
the first $d$ are
\begin{equation}
\ket{\mathcal{B}_\lblvec}
=
\ket{y_\lblvec^{(1)}}
\oplus
\left(\bigoplus_{\lblblk=2}^{k!}\ket{{\zerovec}_d}\right).
\label{eq:first-d-Bj}
\end{equation}
Because all disconnected graphs
are isomorphic by Lemma~\ref{lem:Ablocks}, each of the $k!$ blocks of $d$
eigenvectors is obtained from those in \eqref{eq:first-d-Bj} under a permutation
operator. For $j=(\lblblk-1) d+\lblvec$, with $1\le\lblblk\le k!$ labeling the
block and $1\le\lblvec\le d $ labeling the vector within it,
there exists a $\hat{\pi}_{\lblblk,1}$ such that
$\bra{\mathcal{B}_j}\hat{\pi}_{\lblblk,1}\ket{\mathcal{B}_\beta}=1$.
Therefore one can write the eigenvalue equation for identical hard-core 
bosons on 
paths as $A_{\rm HC}|\mathcal{B}_j\rangle=\lambda_j|\mathcal{B}_j\rangle$, 
where the eigenvectors are 
\begin{equation}
\ket{\mathcal{B}_j}
=
\ket{\mathcal{B}_{(\lblblk-1)d+\lblvec}}
=
\hat{\pi}_{\lblblk,1}\ket{\mathcal{B}_\lblvec}.
\label{eq:bosefunction}
\end{equation}
If the unit antisymmetry operator is defined in the same way as $A_{\rm HC}$,
\begin{equation}
\mathcal{A}\eqdef\bigoplus_{\lblblk=1}^{k!}c_\lblblk I_d,
\label{eq:unit_adj}
\end{equation}
where $I_d$ is the identity matrix of size $d$ and $c_\lblblk\in\{-1,1\}$, then 
Eq.~(\ref{eq:commutativity}) and $[A_{\rm HC},\mathcal{A}]=0$ are both 
satisfied by construction. Thus $\mathcal{A}\hat{\pi}_{\lblblk,1}
|\mathcal{B}_\lblvec\rangle$ is also an eigenstate of the distinguishable
hard-core boson
graph for any choice of the coefficients $c_\lblblk$.

The fermionic wavefunction Eq.~(\ref{eq:fermifunction}) must be completely 
antisymmetric with respect to any permutation of occupation indices. 
Lemma~\ref{lem:Ablocks} states that for hard-core bosons 
on paths each such permutation maps one graph component to another. 
Consequently the wavefunction for distinguishable fermions can always be 
expressed as
\begin{equation}
|\mathcal{F}_j\rangle=\mathcal{A}|\mathcal{B}_j\rangle
=\sigma(p_{\lblblk,1})\hat{\pi}_{\lblblk,1}|\mathcal{B}_\lblvec\rangle,
\label{eq:fermifunction2}
\end{equation}
where $p_{\lblblk,1}$ is the vertex permutation associated with the permutation
operator $\hat{\pi}_{\lblblk,1}$. This corresponds to the application of a unit 
antisymmetry operator $\mathcal{A}$ defined by
$c_\lblblk=\sigma(p_{\lblblk,1})$.
Alternatively, one can interpret the eigenstates of distinguishable hard-core 
bosons (\ref{eq:bosefunction}) as this $\mathcal{A}$ operating on the fermionic 
state defined in Eq.~(\ref{eq:fermifunction}), so that $|\mathcal{B}_j\rangle
=\mathcal{A}|\mathcal{F}_j\rangle$.

\end{prooflem}

Lemma~\ref{lem:fermibose} implies that the $j$th fermionic wavefunction 
(\ref{eq:fermifunction}) or (\ref{eq:fermifunction2}) is always related (via
${\mathcal A}$) to one of the $k!$-degenerate eigenstates of $k$ 
distinguishable hard-core bosons on a path graph (weighted or simple). The 
graph associated with the hard-core boson Hamiltonian is all positive, so the 
maximal eigenvector is also all positive by the Perron-Frobenius theorem (see
Theorem~\ref{thm:PerronFrobenius} for more details); likewise for the ground 
state (eigenstate with most negative eigenvalue) of the many-body Hamiltonian. 
All coefficients of the maximal fermionic wavefunction in a given graph 
component are therefore guaranteed to have the same sign; this is somewhat 
surprising given that the state~(\ref{eq:fermifunction}) is composed of 
(generally signed) single-particle states. 

A second inference that can be drawn from Lemma~\ref{lem:fermibose} is that the 
hard-core boson wavefunctions cannot be derived from symmetrized fermionic 
states for more general graphs. The decomposition of the resulting graph for 
$k$ hard-core bosons into $k!$ graph components is only guaranteed for paths. 
For other graphs, the labels of the deleted vertices will be the same but two 
vertices $v_1=( x_1, \cdots, x_i, x_j,\cdots, x_k )$ and
$v_2=( x_1, \cdots, x_j, x_i,\cdots, x_k )$ that differ by only a single
permutation of labels may still be connected. Thus, there will generically be
fewer than $k!$ graph components, which means that one cannot define $c_\lblblk$
in the unit antisymmetry operator~(\ref{eq:unit_adj}) to transform fermionic 
states to bosonic ones.

An important corollary of Lemma~\ref{lem:fermibose} is that the projections of 
the symmetrized fermionic states into the quotient graph obtained by equitably
partitioning via $\Pi_{\rm I}$ constitute the complete set of eigenvectors for 
identical bosons on paths.

\begin{cor}
The set of projected symmetrized fermionic functions 
$\{Q_{\rm I}^T\mathcal{A}|\mathcal{F}_j\rangle\,|\,1\leq j\leq d\}$,
$d=n!/k!(n-k)!$, constitutes the complete set of eigenvectors for $k$
identical hard-core bosons on the path of length $n$ ($P_n$ or 
$\tilde{P}_n$).
\label{cor:states}
\end{cor}

\begin{proofcor}

All of the eigenvectors for hard-core bosons are necessarily symmetric under
the interchange of particle occupation labels. The equitable partition 
$\Pi_{\rm I}$ groups together all vertices whose labels are equivalent under a 
permutation. Recall that all of the $k!$ graph components of the deleted 
$k$-fold Cartesian powers of $P_n$ (or $\tilde{P}_n$) are isomorphic under
a vertex permutation operator $p$. The normalized partition matrix 
$Q_{\rm I}^T$ therefore projects all eigenvectors that are symmetric under the 
application of the associated permutation operator $\hat{\pi}$ to the 
eigenvector of a single graph component. Likewise, $Q_{\rm I}^T$ maps 
non-symmetric eigenvectors to null vectors, $Q_{\rm I}^T|\mathcal{F}_j\rangle
=|{\zerovec}_d\rangle$. The cardinality of the set of symmetrized fermionic 
wavefunctions of the form~(\ref{eq:bosefermi}) is $d=n!/k!(n-k)!$, and the size 
of each graph component is $d$. Furthermore, 
\begin{eqnarray}
\lambda_jQ_{\rm I}^T|\mathcal{B}_j\rangle
&=&Q_{\rm I}^TDA_{\mathcal{G}^{\Box k}}D|\mathcal{B}_j\rangle\nonumber \\
&=&Q_{\rm I}^TQ_{\rm I}Q_{\rm I}^TDA_{\mathcal{G}^{\Box k}}D
|\mathcal{B}_j\rangle\nonumber \\
&=&Q_{\rm I}^TDA_{\mathcal{G}^{\Box k}}DQ_{\rm I}Q_{\rm I}^T
|\mathcal{B}_j\rangle \nonumber \\
&=&A_{\rm HC,I}Q_{\rm I}^T\mathcal{A}
|\mathcal{F}_j\rangle=\lambda_jQ_{\rm I}^T\mathcal{A}|\mathcal{F}_j\rangle,
\end{eqnarray}
where on the second line we have used $Q_{\rm I}^TQ_{\rm I}=I$ and on the 
third line $[Q_{\rm I}Q_{\rm I}^T,DA_{\mathcal{G}^{\Box k}}D]=0$. Thus, 
the $Q_{\rm I}^T\mathcal{A}|\mathcal{F}_j\rangle$ constitute a complete set of 
eigenvectors for the graph adjacency matrix $A_{\rm HC,I}\eqdef 
Q_{\rm I}^TDA_{\mathcal{G}^{\Box k}}DQ_{\rm I}$, corresponding to $k$ 
identical hard-core bosons on $P_n$ or $\tilde{P}_n$. 

\end{proofcor}

An immediate consequence of Corollary~\ref{cor:states} is that the entire 
spectrum for $k$ identical hard-core bosons on a path graph is given by 
the spectrum of non-interacting distinguishable fermions on the same path.
Of course, this result also follows directly from the TG construction. 
By the ratio condition~(\ref{eq:fin-ratio}), one can also state that symmetric 
powers of hypercubically weighted path graphs are periodic. The following two 
sections address whether these graphs also exhibit PST.

\subsection{Generalized mirror symmetry for hard-core bosons}
\label{sec:Symmetry}

As discussed in Sec.~\ref{sec:PSTnon}, non-interacting bosons (either 
distinguishable or indistinguishable) exhibit PST on weighted path graphs. This 
is because two conditions are satisfied: the eigenvalues of the attendant 
adjacency matrix satisfy the ratio condition~(\ref{eq:ratio}), and the 
eigenvectors of the underlying path graphs are eigenstates of the parity 
operator. Recall that PST occurs between any two vertices of the Cartesian
power graphs (i.e.\ for distinguishable bosons) with labels 
$(i,j,\ldots,k)$ and $(n+1-i,n+1-j,\ldots,n+1-k)$ in time $\tau=\pi/2$.
Suppose for concreteness that one has $k=3$ particles on $\tilde{P}_7$. An 
example of a vertex pair connected via PST would be $v_1=(1,2,3)$ and
$v_2=(7,6,5)$. The labels of $v_1$ are in ascending order while those of 
$v_2$ are in descending order, i.e.\ the ordering of the labels is permuted. 
The particles must pass through each other during the quantum walk, which is 
possible because they are non-interacting. 

\begin{figure}
\begin{tikzpicture}[>= stealth]

	\draw 	(0,0)--(3,0);
	\draw	(5,0)--(8,0);
	\draw	(2.5,1.5)--(5.5,1.5);
	\draw	(2.5,-1.5)--(5.5,-1.5);

	\filldraw[black]	(2,0) circle (4pt)
						(3,0) circle (4pt)
						(5,0) circle (4pt)
						(6,0) circle (4pt)
						(4.5,1.5) circle (4pt)
						(5.5,1.5) circle (4pt)
						(2.5,-1.5) circle (4pt)
						(3.5,-1.5) circle (4pt);
						
	\filldraw[red]		(1,0) circle (4pt)
						(2.5,1.5) circle (4pt)
						(8,0) circle (4pt)
						(4.5,-1.5) circle (4pt);
						
	\filldraw[yellow]	(0,0) circle (4pt)
						(3.5,1.5) circle (4pt)
						(7,0) circle (4pt)
						(5.5,-1.5) circle (4pt);
		
				\draw[<->] (2,1.5) ..controls (1,1.5) and (1,1)..(1,0.5);
				\draw[<->] (6,1.5) ..controls (7,1.5) and (7,1)..(7,0.5);		
				\draw[<->] (1,-0.5) ..controls (1,-1) and (1,-1.5)..(2,-1.5);
				\draw[<->] (7,-0.5) ..controls (7,-1) and (7,-1.5)..(6,-1.5);
				\draw[<->,very thick,blue!50!white] (4,1.25)--
(4,-1.25);
				
		\draw (0.5,1) node{\large$\hat{\pi}_{\perp}$};
		\draw (7.5,-1) node{\large$\hat{\pi}_{\perp}$};
		\draw (7.5,1) node{\large$\hat{P}$};
		\draw (0.5,-1) node{\large$\hat{P}$};
		\draw (4.25,0.25) node{\large$\hat{C}$};

\end{tikzpicture}
\caption{\small Occupation space representation of the composite 
$\hat{C}$ operator on a given configuration space site.  The central 
arrow shows the composite operation directly as a whole while the side 
configurations exhibit the constituent operations, explicitly showing 
the commutativity of $\hat{\pi}_{\perp}$ and $\hat{P}$.}
\label{sigma-C_comm}
\end{figure}
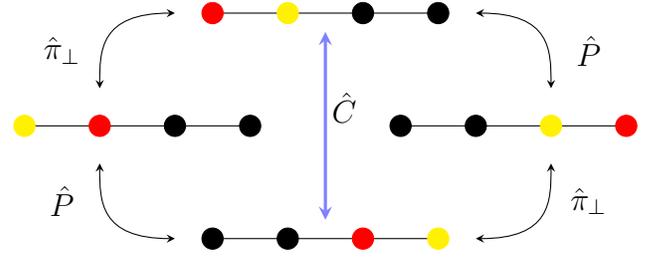

Hard-core bosons, in contrast, cannot pass through each other during the 
quantum walk. The initial ordering of the vertex labels must be preserved 
for any pair of vertices connected via PST. In the previous example for instance, 
vertices $v_1$ and $v_2$ are located in different graph components and are 
disconnected. Rather, hard-core bosons require PST-connected pairs of vertices
$(i,j,k)$ and $(p,q,r)$ where $i<j<k$ and $p<q<r$. Thus, satisfying the 
ratio condition~(\ref{eq:ratio}) and ensuring the eigenstates of the underlying 
graph are eigenfunctions of the parity operator is not sufficient to ensure PST
with hard-core bosons.

Consider the composite operator $\hat{C}$,
\begin{equation}
\hat{C}\eqdef \hat{\pi}_{\perp}\hat{P},
\label{eq:C-operator}
\end{equation}
defined as the product of the global parity operator $\hat{P}$ and the 
permutation operator $\hat{\pi}_\perp(v)$ effecting the permutation 
$p_{\perp}$ of 
the list corresponding to vertex label $v$ to yield its mirror-symmetric 
ordering. For example, for a vertex $v=(1,3,5)$ with three distinguishable bosons, 
$p_{\perp}(1,3,5) = (5,3,1)$. The mapping of vertex $v_1$ to $v_2$ via
$\hat{C}$ can be considered as a reflection from $v_1$ to $v_2$ across a
generalized mirror plane $M$. The $\hat{\pi}_{\perp}$ operator ensures that 
the ordering of vertex labeling is restored after having been reversed under 
the application of $\hat{P}$. Figure~\ref{sigma-C_comm} shows the operation of 
$\hat{C}$ in occupation space on a state of the system consisting of two
hard-core bosons on $\tilde{P_4}$. The permutation operator and global parity 
operator commute, which follows directly from the fact that the sign of the 
permutation, $\sigma[p_k(v)]$ only depends on the number of bosons in the 
system and is independent of any other operation. The effect of 
$\hat{\pi}_{\perp}$ on a given state is independent of the spatial state, while 
the parity operates only on the spatial degrees of freedom.

\begin{figure}
\begin{tikzpicture}

	\draw[blue!50!white,very thick]	(3,4.5).. controls (3.125, 3.75)
and (3.75, 3.125)..(4.5,3);
	\draw[blue!50!white,very thick]	(1.5,4.5).. controls (1.75,3) 
and (3,1.75)..(4.5,1.5);
	\draw[blue!50!white,very thick]	(0,3).. controls (1.25,2.75) 
and (2.75,1.25)..(3,0);
	\draw[blue!50!white,very thick]	(0,1.5).. controls (0.75, 1.375)
and (1.375, 0.75)..(1.5,0);

	\filldraw[blue!50!white] (3,4.5) 	circle (0.25);
	\filldraw[blue!50!white] (4.5,3) 	circle (0.25);
	\filldraw[blue!50!white] (1.5,4.5) 	circle (0.25);
	\filldraw[blue!50!white] (4.5,1.5) 	circle (0.25);
	\filldraw[blue!50!white] (0,3) 		circle (0.25);
	\filldraw[blue!50!white] (3,0) 		circle (0.25);
	\filldraw[blue!50!white] (0,1.5) 	circle (0.25);
	\filldraw[blue!50!white] (1.5,0) 	circle (0.25);
	\filldraw[blue!50!white] (0,0) 		circle (0.25);
	\filldraw[blue!50!white] (1.5,1.5) 	circle (0.25);
	\filldraw[blue!50!white] (3,3) 		circle (0.25);
	\filldraw[blue!50!white] (4.5,4.5) 	circle (0.25);

	\filldraw[black]	(0,1.5) 	circle (4pt)
						(0,3) 		circle (4pt)						
						(4.5,4.5) 	circle (4pt)
						(1.5,0) 	circle (4pt)
						(3,3) 		circle (4pt)
						(1.5,4.5) 	circle (4pt)
						(3,0) 		circle (4pt)
						(1.5,1.5) 	circle (4pt)
						(3,4.5) 	circle (4pt)
						(0,0) 		circle (4pt)
						(4.5,1.5) 	circle (4pt)
						(4.5,3) 	circle (4pt)
						
						(6,4.5)		circle (4pt)
						(7,4.5)		circle (4pt)
						(8,4.5)		circle (4pt)
						(7,3.5)		circle (4pt)
						(6,0)		circle (4pt)
						(7,0)		circle (4pt)
						(8,0)		circle (4pt)
						(7,1)		circle (4pt);
						
						\draw	(2,2.5)	node{$M$};

	\draw (-0.25,2.25) node{$2$};
	\draw (-0.25,0.75) node{$\sqrt{3}$};
	\draw (0.75,-0.25) node{$\sqrt{3}$};
	\draw (0.75,1.75) node{$\sqrt{3}$};
	\draw (1.75,0.75) node{$\sqrt{3}$};
	\draw (2.25,-0.25) node{$2$};

	\draw (4.75,2.25) node{$2$};
	\draw (4.75,3.75) node{$\sqrt{3}$};
	\draw (3.75,4.75) node{$\sqrt{3}$};
	\draw (3.75,2.75) node{$\sqrt{3}$};
	\draw (2.75,3.75) node{$\sqrt{3}$};
	\draw (2.25,4.75) node{$2$};
	
	\draw (6.5,4.75)  node{$2$};
	\draw (7.5,4.75)  node{$\sqrt{6}$};
	\draw (6.75,4)    node{$\sqrt{6}$};

	\draw (6.5,-0.25)  node{$2$};
	\draw (7.5,-0.25)  node{$\sqrt{6}$};
	\draw (6.75,0.5)   node{$\sqrt{6}$};
	
	\draw (2.25,-1) node{(a)};
	\draw (7,-1)    node{(b)};

	\draw 	(0,0)--(0,3);
	\draw	(1.5,0)--(1.5,1.5);
	\draw	(4.5,1.5)--(4.5,4.5);
	\draw	(0,0)--(3,0);
	\draw	(3,3)--(3,4.5);
	\draw	(0,1.5)--(1.5,1.5);
	\draw	(3,3)--(4.5,3);
	\draw	(1.5,4.5)--(4.5,4.5);
	\draw[red, thick, dashed] (0,0)--(4.5,4.5);
	
	\draw	(6,4.5)--(8,4.5);
	\draw	(7,4.5)--(7,3.5);
	\draw	(6,0)--(8,0);
	\draw	(7,0)--(7,1);
	
\end{tikzpicture}
\caption{\small Configuration space graph of two hard-core bosons on 
$\tilde{P_4}$. (a) The mirror plane $M$ is shown as the dashed line on the 
anti-diagonal.  Vertices not in singleton cells of the partition
are grouped in pairs and connected by the curved blue lines. The cells are symmetrically disposed 
about the anti-diagonal line. (b) The resulting quotient graph.}
\label{hard-core-eff_2}
\end{figure}
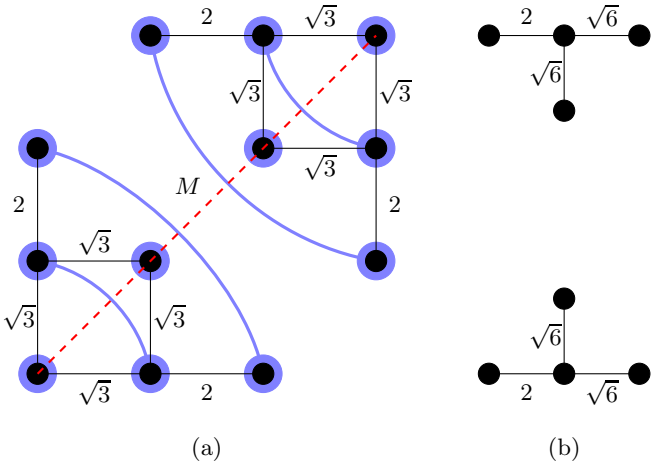

Consider the generalized equitable partition $\Pi_C$ where cells are comprised 
of all vertices that are equivalent under $\hat{C}$. The associated normalized
partition matrix is defined as
\begin{equation}
(Q_{C}^T)_{ij} = 
\begin{cases}
\Omega(v_j) & \text{if vertex $v_j$ is in cell $C_i$}\\
0 & \text{otherwise}
\end{cases}.
\end{equation}
Figure~\ref{hard-core-eff_2} depicts the case of two distinguishable hard-core 
bosons on $\tilde{P}_4$. The mirror plane $M$ (shown as a dashed red line) and 
the associated equitable partitioning $\Pi_C$ (blue outlines and curved lines) 
are shown in Fig.~\ref{hard-core-eff_2}(a). Figure~\ref{hard-core-eff_2}(b)
shows the two graph components of the resulting quotient graph. Suppose the 
$i$th eigenvector $|Z_i\rangle$ of the hard-core graph is a 
simultaneous eigenvector of $\hat{C}$ with eigenvalues $\pm 1$,
$\hat{C}|Z_i\rangle = \pm|Z_i\rangle$. Then 
$Q_C^T|Z_i\rangle =|\zerovec\rangle$ for all antisymmetric $|Z_i\rangle$ under
$\hat{C}$, that is for values of $i$ where $\hat{C}|Z_i\rangle = -|Z_i\rangle$.
Following the same arguments as found in Sec.~\ref{sec:PSTnon}, vertex 
periodicity in the quotient graph under $\Pi_C$ will be a direct manifestation 
of PST in the original hard-core boson graph. The remainder of this section is 
devoted to proving that this is indeed the case.

\begin{lemma}
Consider a vertex state $|v\rangle$ in one of the $k!$ graph components of a
system of $k$ hard-core bosons on $P_n$ or $\tilde{P}_n$. Given its expansion 
in terms of the 
$d$ orthogonal hard-core boson eigenstates with support on the component
$|v\rangle = \sum_{i}\alpha_i|\mathcal{B}_i\rangle$, the vertex state 
$|u\rangle = \hat{C}|v\rangle$ has the same expansion up to sign 
differences in the expansion coefficients, $\alpha\to\pm\alpha$.
\label{lem:expansion}
\end{lemma}

\begin{prooflem}
Let the set $\{|\mathcal{B}_i\rangle\}$ denote the orthonormal eigenvectors 
of $k$ distinguishable hard-core bosons on $P_n$ or $\tilde{P}_n$, given in 
Lemma~\ref{lem:fermibose} as 
$|\mathcal{B}_i\rangle=\mathcal{A}|\mathcal{F}_i\rangle$.
One can express a vertex state $|v\rangle$ on the first of the $k!$ isomorphic
graph components in terms of this eigenbasis:
\begin{equation}
|v\rangle = \sum\limits_{i=1}^{k!d}|\mathcal{B}_i\rangle\langle\mathcal{B}_i|v
\rangle=\sum\limits_{i=1}^{d}\alpha_i|\mathcal{B}_i\rangle,
\label{eq:vexpansion}
\end{equation}
where $\alpha_i\eqdef\langle\mathcal{B}_i|v\rangle$ for
$1\le i\le d$, and $\langle\mathcal{B}_i|v\rangle=0$ for $d<i\le k!d$.
That is, because the eigenvectors 
can be decomposed into $k!$ cardinality-$d$ contributions according to 
Eq.~(\ref{eq:bosefunction}), the sum over $k!d$ terms can be restricted to 
only those $d$ 
terms that have non-zero overlap with $|v\rangle$. Under the operation of 
$\hat{C}$, any vertex is mapped to its mirror-symmetric (about the plane $M$)
state 
\begin{eqnarray}
|u\rangle\eqdef\hat{C}|v\rangle&=&\sum\limits_{i=1}^{d}\alpha_i
\hat{\pi}_{\perp}\hat{P}\mathcal{A}|\mathcal{F}_i\rangle\nonumber \\
&=&\sum\limits_{i=1}^{d}\alpha_i\mathcal{A}
\sum\limits_{p\in S_k}\sigma\left[
p\left(\ell_i\right)\hat{\pi}_{\perp}\right]
\hat{\pi}_{\perp}\bigotimes_{j=1}^k
\hat{P}|z_{\ell_i(j)}\rangle,
\nonumber
\end{eqnarray}
where the fermionic functions are given by Eq.~(\ref{eq:fermifunction}) in
terms of the eigenstates of the path graphs $|z_j\rangle$ 
($|\mathcal{K}_j\rangle$ for hypercubically weighted path graphs). The result 
makes use of the fact that 
$[\mathcal{A},\hat{C}]=0$, which follows from the fact that $\hat{C}$ maps a 
vertex to another in the same graph component and $\mathcal{A}$ is 
block-diagonal in graph components, c.f.\ Eq.~(\ref{eq:Ablockdiag}). Note also
that we have included the sign of the effect of the $\hat{\pi}_{\perp}$ in the
sign operator $\sigma$ because $\hat{\pi}_{\perp}$ will exchange fermionic
particles. As discussed in Sec.~\ref{sec:PSTnon}, the parity operator on each 
path graph
eigenvector returns $\hat{P}|z_j\rangle=(-1)^j|z_j\rangle$. Suppose that the
fermionic state $|\mathcal{F}_i\rangle$ has $n_e(i)$ particles in even-$j$ 
states and $n_o(i)$ particles in odd-$j$ states, where $n_e(i)+n_o(i)=k$ for 
all $i$; that is, there are $n_o(i)$ odd integers and $n_e(i)$ even integers in
$\ell_i$. Then $\hat{P}$ on the full state returns $(-1)^{n_o(i)}$:
\begin{equation}
|u\rangle
=\sum\limits_{i=1}^{d}\alpha_i(-1)^{n_o(i)}\mathcal{A}\sum\limits_{p\in S_k}
\sigma\left[p\left(\ell_i\right)\hat{\pi}_{\perp}\right]\hat{\pi}_{\perp}
\bigotimes_{j=1}^k|z_{\ell_i(j)}\rangle,
\nonumber
\end{equation}

Meanwhile, $\hat{\pi}_{\perp}$ completely reverses the particle ordering. This
is equivalent to 
\begin{eqnarray}
\hat{\pi}_{\perp}\bigotimes_{j=1}^k|z_{\ell_i(j)}\rangle
&=&|z_{\ell_i(1)}\rangle\otimes|z_{\ell_i(2)}\rangle\cdots
\otimes|z_{\ell_i(k)}\rangle\nonumber \\
&=&|z_{\ell_i(k)}\rangle\otimes 
|z_{\ell_i(k-1)}\rangle\cdots\otimes|z_{\ell_i(1)}\rangle.\quad
\end{eqnarray}
This is a completely antisymmetric permutation of the original ordering. 
Whether this changes the sign of the coefficient depends on the number of 
particles. If $k$ is even, then $p\left(\hat{\pi}_{\perp}\right)=(-1)^{k/2}$
while for odd $k$ then $p\left(\hat{\pi}_{\perp}\right)=(-1)^{(k-1)/2}$; more
concisely $p\left(\hat{\pi}_{\perp}\right)=(-1)^{k\setminus 2}$. The result is
\begin{eqnarray}
|u\rangle
&=&\sum\limits_{i=1}^{d}\alpha_i(-1)^{n_o(i)+k\setminus 2}\mathcal{A}
\sum\limits_{p\in S_k}\sigma\left[p\left(\ell_i\right)\hat{\pi}_{\perp}\right]
\bigotimes_{j=1}^k|z_{\ell_i(j)}\rangle\nonumber \\
&=&\sum_{i=1}^d\alpha_i(-1)^{n_o(i)+k\setminus 2}|\mathcal{B}_i\rangle.
\label{eq:uexpansion}
\end{eqnarray}
The expansion coefficients $\alpha_i'=\alpha_i(-1)^{n_o(i)+k\setminus 2}$ for 
the state $|u\rangle=\hat{C}|v\rangle$ are therefore equal to the expansion 
coefficients $\alpha_i$ for $|v\rangle$, up to a sign depending on $n_o(i)$ and
the number of particles $k$. 
\end{prooflem}

One can relate the sign of the expansion coefficients in 
Eq.~(\ref{eq:uexpansion}) to the eigenvalues of $k$ hard-core bosons on the 
hypercubically weighted path graph $\tilde{P}_n$. Recall that the eigenvalues 
of $\tilde{P}_n$ are $\tilde{\lambda}=-(n-1)+2\{0,1,\ldots,n-1\}$. The 
eigenvalues for $k$ identical hard-core bosons on $\tilde{P}_n$ are therefore 
$\lambda_i=-k(n-1)+2\sum_j\ell_i(j)$. Because one can factor out a common
$-k(n-1)$ term, the eigenvalues are determined by a positive integer 
$a\eqdef\sum_j\ell_i(j)$, the sum of $k$ unique non-negative integers (the 
unique compositions) corresponding to the elements of the site occupations 
$\ell_i$. Suppose that $a$ is even. If $k=2$ then there are $(a/2)+1$ 
different choices for the $\ell_i$, each of which contains either two odd or 
two even integers; in both cases one obtains $(-1)^{n_o}=1$. For $k=3$, one
of the two even (odd) integers above can be partitioned into the sum of two 
smaller integers, which will be both even or odd (one even and one odd); the 
total number of odd integers in the $k=3$ composition will be even and again 
$(-1)^{n_o}=1$. This argument can be extended to arbitrarily many partitions 
of the integers comprising the composition, subject to the uniqueness 
constraint. A similar argument yields that for odd $a$ one has 
$(-1)^{n_o}=-1$. Thus, if the eigenvalue of $|{\mathcal B}_i\rangle$ is
$\lambda_i=-k(n-1)+2a$ with $a$ a non-negative integer, then the expansion 
coefficients in Eq.~(\ref{eq:uexpansion}) can be written 
$\alpha_i(-1)^{a+k\setminus 2}$. 

The maximum eigenvalue is given by
\begin{eqnarray}
\lambda_{\rm max}&=&-k(n-1)+2\sum_{j=n-k}^{n-1}j\nonumber \\
&=&-k(n-1)+k(2n-k-1)=k(n-k),
\end{eqnarray}
so the $k$-boson eigenvalues are 
\begin{equation}
\lambda_i=k(k-n)+2(i-1),\quad i\in\{1,2,\ldots,k(n-k)\}.
\label{eq:eigenvalues}
\end{equation}
The minimum eigenvalue for $k$ hard-core bosons on $\tilde{P}_n$ is
\begin{eqnarray}
\lambda_{\rm min}&=&-k(n-1)+2\sum_{j=0}^{k-1}j\nonumber \\
&=&-k(n-1)+k(k-1)=k(k-n),
\end{eqnarray}
so that $a_{\rm min}=k(k-1)/2$. One can make use of the fact that 
$(-1)^{k(k-1)/2}=(-1)^{k\setminus 2}$ for all $k$, so that 
$(-1)^{a_{\rm min}+k\setminus 2}=1$. The first expansion coefficient, the
overlap of the site $u$ with the lowest bosonic eigenstate
$\langle{\mathcal B}_1|u\rangle=\alpha_1(-1)^{n_o(1)+k\setminus 2}$, therefore 
coincides with $\alpha_1=\langle{\mathcal B}_1|v\rangle$. For the second 
expansion coefficient, the only 
quantity that can change is $a$. The second-lowest eigenvalue corresponds to a
single particle in a higher single-particle state, i.e.\ an energy $k(k-n)+2$.
If $a$ increases by unity, then by the argument above $(-1)^{n_0}$ changes sign
relative to that for the lowest state; necessarily then 
$\alpha_2(-1)^{n_o(2)+k\setminus 2}=-\alpha_2$ for $|u\rangle$ while it is 
$\alpha_2$ for $|v\rangle$. Because $a$ is always composed of sums of integers,
the expansion coefficient for $|u\rangle$ will change sign at each successive
eigenvalue.

The remaining eigenvalues $\lambda_i$ are degenerate,
corresponding to common values of 
$a_i=\sum_m\ell_i(m)$. Defining the set $\overline{a}_i$ with elements 
$\ell_i$ such that $\sum_j\ell_i(j)=a_i$, and
the vectors $|\overline{\mathcal{B}}_i^{(j)}\rangle$ corresponding
to each degenerate $|{\mathcal B}_i\rangle$ eigenvector,
one can formalize this observation by defining the spectral decomposition of
the distinguishable hard-core adjacency matrix as
\begin{equation}
A_{\rm HC}=\sum_{i=1}^{k(n-k)}\lambda_i\sum_{j=1}^{|\overline{a}_i|}
|\overline{\mathcal B}_i^{(j)}\rangle\langle\overline{\mathcal B}_i^{(j)}|,
\label{eq:specdecompalt}
\end{equation}
This allows one to rewrite Eq.~(\ref{eq:vexpansion}) as
\begin{equation}
|v\rangle=\sum_{i=1}^{k(n-k)}\sum_{j=1}^{|\overline{a}_i|}
\overline{\alpha}_i^{(j)}|\overline{\mathcal B}_i^{(j)}\rangle,
\label{eq:vexpandalt}
\end{equation}
where $\overline{\alpha}_i^{(j)}\eqdef\langle\overline{\mathcal B}_i^{(j)}
|v\rangle$, and Eq.~(\ref{eq:uexpansion}) as
\begin{equation}
|u\rangle=\hat{C}|v\rangle=\sum_{i=1}^{k(n-k)}(-1)^{i-1}
\sum_{j=1}^{|\overline{a}_i|}\overline{\alpha}_i^{(j)}
|\overline{\mathcal B}_i^{(j)}\rangle.
\label{eq:uexpandalt}
\end{equation}
These alternate expressions readily yield the following lemma.

\begin{lemma}

Equitable partitioning of the configuration space graph representing $k$ 
hard-core bosons on $\tilde{P}_n$ with respect to the operator $\hat{C}$ in 
each disconnected graph component will preserve the lowest eigenvector but 
annihilate every second higher eigenvector, in all degeneracy, from the 
distinguishable spectrum. 
\label{lem:sigma-spec}
\end{lemma}

\begin{prooflem}

Consider the quotient graph $\Gamma_C\eqdef\left(\tilde{P}_n^{\Box k}
\setminus D\right)/\Pi_C$, where $\Pi_C$ is an equitable partitioning about
$\hat{C}$. The associated normalized partition matrix $Q_C$ groups vertices
connected by $\hat{C}$ into cells. Because $Q_C$ is even, the vertex states
$|v'\rangle$ in $A_{\Gamma_C}$ are superpositions of the $|v\rangle$ and 
$|u\rangle$ above:
\begin{eqnarray}
|v'\rangle&=&\frac{1}{\sqrt{2}}Q_C^T\left(|v\rangle+\hat{C}|v\rangle\right)
\nonumber \\
&=&\sum_{i=1}^{k(n-k)}\frac{1}{\sqrt{2}}\left[1+(-1)^{i-1}\right]
\sum_{j=1}^{|\overline{a}_i|}\overline{\alpha}_i^{(j)}Q_C^T
|\overline{\mathcal B}_i^{(j)}\rangle\nonumber \\
&=&\operatorname*{\sum{}^{\prime}}_{i=1}^{k(n-k)}\sum_{j=1}^{|\overline{a}_i|}
\overline{\alpha}_i^{(j)}|\overline{B}_i^{(j)}\rangle,
\end{eqnarray}
where $|\overline{B}_i\rangle\eqdef Q_C^T|\overline{\mathcal B}_i\rangle$
and in the second line we have used Eqs.~(\ref{eq:vexpandalt}) and 
(\ref{eq:uexpandalt}). The effective expansion coefficient is zero for all even 
$i$ in the above sum,
corresponding to every second (degenerate) eigenvalue. All odd-$i$ terms are
preserved, including the lowest eigenstate. The primed sum
in the last line (i.e.\ ``$\,\sum{}^{\prime}$'')
denotes the absence of even-$i$ terms in the sum. Because the eigenvector for 
every second eigenstate is missing from the expansion of an arbitrary vertex 
state in the quotient graph, the spectral decomposition of the quotient 
adjacency matrix can be written
\begin{eqnarray}
Q_C^TA_{\rm HC}Q_C
&=&
\operatornamewithlimits{\sum{}^{\prime}}_{i=1}^{k(n-k)}
\lambda_i
\sum_{j=1}^{|\overline{a}_i|}Q_C^T|\overline{\mathcal B}_i^{(j)}\rangle
\langle\overline{\mathcal B}_i^{(j)}|Q_C\nonumber \\
&=&\sum_{i=1}^{k(n-k)/2}\tilde{\lambda}_i\sum_{j=1}^{|\overline{a}_i|}
|\overline{B}_i^{(j)}\rangle\langle\overline{B}_i^{(j)}|.
\label{eq:HCunderC}
\end{eqnarray}
Thus, while the eigenvalues of $A_{\rm HC}$ are $\lambda_i=k(k-n)+2(i-1)$, 
$i=\{1,2,\ldots,k(n-k)\}$, the eigenvalues of $Q_C^TA_{\rm HC}Q_C$ are
$\tilde{\lambda}_i=k(k-n)+4(i-1)$, $i=\{1,2,\ldots,k(n-k)/2\}$. 

\end{prooflem}

Recall from the proof of Corollary~\ref{cor:states} that the 
indistinguishability equitable partition $Q_{I}$ is a $k!$-to-one map that 
groups all vertices with indices belonging to the same symmetric group into
one cell in the quotient graph. The eigenvalues $\tilde{\lambda}_i$ in 
Eq.~(\ref{eq:HCunderC}) are then $k!$ degenerate as a result of 
Lemma~\ref{lem:Ablocks}. The effect of applying $Q_{I}$ will therefore be to
eliminate this degeneracy in the spectrum of the quotient. It follows that
the adjacency matrix
for identical hard-core bosons under $\Pi_C$ will be
\begin{equation}
Q_C^TA_{\rm HC,I}Q_C=\sum_{i=1}^{k(n-k)/2}\tilde{\lambda}_i
\sum_{j=1}^{|\overline{a}_i|}
Q_C^T|\tilde{\mathcal B}_i^{(j)}\rangle\langle\tilde{\mathcal B}_i^{(j)}|Q_C,
\label{eq:QCAHCIQC}
\end{equation}
where $|\tilde{\mathcal B}_i^{(j)}\rangle
\eqdef Q_I^T|\overline{\mathcal B}_i^{(j)}\rangle$, $A_{\rm HC,I}=Q_I^T
A_{\rm HC}Q_T$, and $\tilde{\lambda}_i$ are eigenvalues defined in 
Lemma~\ref{lem:sigma-spec}. Note that the eigenvalue
degeneracy corresponding to the cardinality of the set $\overline{a}_i$ is
unchanged under $\Pi_C$: the number of linearly independent eigenvectors is 
of order $d/2$, while the number of unique eigenvalues is the (generally lower)
value $k(n-k)/2$. The increase in the eigenvalue spacing from 2 to 4 
in the quotient graph under the equitable partitioning $\Pi_C$ will have 
important consequences for PST in identical hard-core bosons, as 
discussed in the next section.

\subsection{PST for hard-core bosons on weighted path graphs}

This section presents the primary result of this work, that hard-core bosons on 
the weighted path graph $\tilde{P}_n$ undergo perfect quantum state transfer. 
The result are proven in two ways. The first proof makes direct use of the time 
evolution operator, while the second infers the presence of PST from the vertex
periodicity of the quotient graph under $\Pi_C$. Before stating the proofs,
it is useful to introduce the following lemma.
 
\begin{lemma}
Vertices in the configuration-space representation of $k$ hard-core 
identical bosons on $\tilde{P}_n$ are all periodic in time $t=\pi$.
\label{lem:period}
\end{lemma}

\begin{prooflem}

The eigenvalues for $k$ bosons on the weighted path graph $\tilde{P}_n$ are 
given by Eq.~(\ref{eq:eigenvalues}). The spectral decomposition of the 
associated adjacency matrix immediately yields the time evolution operator
\begin{eqnarray}
U_{\rm HC,I}(t)&=&\sum_{j=1}^{k(n-k)}e^{-it\lambda_j}
\sum_{m=1}^{|\overline{a}_j|}|\tilde{\mathcal B}_j^{(m)}\rangle
\langle\tilde{\mathcal B}_j^{(m)}|\nonumber \\
&=&e^{-it[k(k-n)-2]}\sum_{j=1}^{k(n-k)}e^{-2itj}\sum_{m=1}^{|\overline{a}_j|}
|\tilde{\mathcal B}_j^{(m)}\rangle\langle\tilde{\mathcal B}_j^{(m)}|.\nonumber
\end{eqnarray}
If $t=\pi$ then the sums over $j$ and $m$ constitute a resolution of the 
identity and 
\begin{equation}
U_{\rm HC,I}(\pi)=e^{-i\pi k(k-n)}I_d.
\end{equation}
Therefore every vertex in the configuration space representation of $k$ 
identical hard-core bosons on $\tilde{P}_n$ is periodic in time $t=\pi$.
\end{prooflem}

\subsubsection{PST by application of the time evolution operator}

\begin{theorem}
A system of $k$ identical hard-core bosons on the weighted path
$\tilde{P}_n$ undergoes PST in time $t=\pi/2$ between two 
configuration-space vertex states
connected by the operation of $\hat{C} = \pi_{\perp}\hat{P}$.
\label{thm:PST1}
\end{theorem}

\begin{proofthm}[PST of Hardcore Bosons I]

PST between vertex states $|v\rangle$ and $|u\rangle$ in the configuration 
space of identical hard-core bosons is defined by Eq.~\eqref{eq:PST}.
One can use Eqs.~(\ref{eq:partition}) and (\ref{eq:vunderQ}) to write the
criterion for PST as
\begin{eqnarray}
\langle u|U_{\rm HC}|v\rangle&=&\langle u|Q_IQ_I^TU_{\rm HC}Q_IQ_I^T|v\rangle
\nonumber \\
&=&\langle\tilde{u}|U_{\rm HC,I}(t)|\tilde{v}\rangle = \gamma,
\label{eq:thm1eq1}
\end{eqnarray}
where $|\gamma|=1$; the above makes use of the first of the two 
properties~(\ref{eq:q-prop1}). From Lemma~\ref{lem:period}, all vertices are periodic in 
time $t=\pi$, so if PST occurs between two vertices then the PST time must be
an even fraction of $\pi$, i.e.\ $\pi/2n$ for some positive integer $n$.

Consider $t_P=\pi/2$. 
Making use of Lemma~\ref{lem:expansion} and Eqs.~(\ref{eq:vexpandalt}) and
(\ref{eq:uexpandalt}), the expanded form of the above equation can be expressed 
as
\begin{widetext}
\begin{eqnarray}
\langle\tilde{u}|U_{\rm HC,I}(t)|\tilde{v}\rangle
&=&\sum_{l=1}^{k(n-k)}(-1)^{l-1}\sum_{j=1}^{|\overline{a}_i|}
\overline{\alpha}_i^{(j)}\langle\tilde{\mathcal B}_i^{(j)}|
\sum_{j''=1}^{k(n-k)}e^{-2itj''}\sum_{m=1}^{|\overline{a}_{j''}|}|
\tilde{\mathcal B}_{j''}^{(m)}\rangle\langle\tilde{\mathcal B}_{j''}^{(m)}|
\sum_{l'=1}^{k(n-k)}\sum_{j'=1}^{|\overline{a}_{l'}|}
\overline{\alpha}_{l'}^{(j')}|\tilde{\mathcal B}_{l'}^{(j')}\rangle
e^{-it[k(k-n)-2]}\nonumber \\
&=&e^{-it[k(k-n)-2]}\sum_{l=1}^{k(n-k)}(-1)^{l-1}e^{-2itl}
\sum_{j'=1}^{|\overline{a}_{l}|}\left(\overline{\alpha}_{l}^{(j')}\right)^2.
\end{eqnarray}
\end{widetext}
If $t=t_P=\pi/2$ then $e^{-2it_Pl}=(-1)^{l}$, and one obtains
\begin{equation}
\langle\tilde{u}|U_{\rm HC,I}(t_P)|\tilde{v}\rangle=\gamma
\sum_{l=1}^{k(n-k)}\sum_{j'=1}^{|\overline{a}_{l}|}\left(
\overline{\alpha}_{l}^{(j')}\right)^2,
\label{eq:thm1eq2}
\end{equation}
where $\gamma=e^{-i\pi k(k-n)/2}$. The summation is the resolution of unity, 
reflecting the orthonormality of the bosonic eigenvectors. Consider the last 
term:
\begin{eqnarray}
\left(\overline{\alpha}_i^{(j)}\right)^2&=&\langle v
|\overline{\mathcal B}_i^{(j)}\rangle\langle\overline{\mathcal B}_i^{(j)}
|v\rangle=\langle\tilde{v}|\tilde{\mathcal B}_i^{(j)}\rangle
\langle\tilde{\mathcal B}_i^{(j)}|\tilde{v}\rangle
\end{eqnarray}
Inserting into Eq.~(\ref{eq:thm1eq2}) gives
\begin{eqnarray}
\langle\tilde{u}|U_{\rm HC,I}(t_P)|\tilde{v}\rangle&=&\gamma\langle
\tilde{v}|\sum_{l=1}^{k(n-k)}\sum_{j'=1}^{|\overline{a}_{l}|}
|\tilde{\mathcal B}_i^{(j)}\rangle\langle\tilde{\mathcal B}_i^{(j)}
|\tilde{v}\rangle\nonumber \\
&=&e^{-i\pi k(k-n)/2},
\end{eqnarray}
so that PST between $|\tilde{v}\rangle$ and $|\tilde{u}\rangle
=\hat{C}|\tilde{v}\rangle$ occurs in time $t_P=\pi/2$ yielding a phase 
$\gamma=e^{-i\pi k(k-n)/2}$.

\end{proofthm}

\subsubsection{PST by vertex periodicity in quotient graph under $\Pi_C$}

\begin{lemma}
The quotient graph resulting from the application of the
equitable partition $\Pi_C$ to the
graph corresponding to $k$ identical hard-core bosons is periodic in 
time $t_P=\pi/2$.
\label{lem:config-period}
\end{lemma}

\begin{prooflem}

The proof of this Lemma follows that of Lemma~\ref{lem:period}, with some minor
modifications. The time evolution operator after equitably partitioning the 
graph for $k$ identical hard-core bosons under $\Pi_C$ is 
\begin{eqnarray}
Q_C^TU_{\rm HC,I}(t)Q_C&=&\sum_{j=1}^{k(n-k)/2}e^{-it\tilde{\lambda}_j}
\sum_{m=1}^{|\overline{a}_i|}Q_C^T|\tilde{\mathcal B}_j^{(m)}\rangle
\langle\tilde{\mathcal B}_j^{(m)}|Q_C\nonumber \\
&=&e^{-it[k(k-n)-4]}\sum_{j=1}^{k(n-k)/2}e^{-4itj}\nonumber \\
&\times&Q_C^T\left(\sum_{m=1}^{|\overline{a}_j|}|\tilde{\mathcal B}_j^{(m)}
\rangle\langle\tilde{\mathcal B}_j^{(m)}|\right)Q_C,\nonumber
\end{eqnarray}
where the eigenvalues $\tilde{\lambda}_j$ are given in 
Lemma~\ref{lem:sigma-spec}. If $t_P=\pi/2$ then the sums over $j$ and $m$ 
constitute a resolution of the identity and 
\begin{equation}
Q_C^TU_{\rm HC,I}(\pi/2)Q_C=e^{-i\pi k(k-n)/2}I_{d'},
\end{equation}
where $d'\approx d/2$. Therefore every vertex in the quotient graph of $k$ 
identical hard-core bosons on $\tilde{P}_n$ under $\Pi_C$ is periodic in time 
$t_P=\pi/2$.

\end{prooflem}

One can now employ a modified version of the discussion in 
Sec.~\ref{sec:PSTnon} to prove that vertex periodicity in the quotient graph 
discussed in Lemma~\ref{lem:config-period} implies PST for hard-core 
identical bosons.

\begin{theorem}
A system of $k$ identical hard-core bosons on the weighted path
$\tilde{P}_n$ undergoes PST in time $t_p=\pi/2$ between two vertex states
connected by the operation of $\hat{C} = \pi_{\perp}\hat{P}$ if the graph 
quotient under the equitable partition $\Pi_C$ is periodic in time $t_p=\pi/2$.
\label{thm:PST2}
\end{theorem}

\begin{proofthm}[PST of Hardcore Bosons II]

Following Theorem~\ref{thm:PST1}, one can write
\begin{eqnarray}
\langle u|U_{\rm HC}(t)|v\rangle&=&\langle\tilde{u}|U_{\rm HC,I}(t)|\tilde{v}\rangle\nonumber \\
&=&\langle\tilde{u}|Q_CQ_C^TU_{\rm HC,I}(t)Q_CQ_C^T|\tilde{v}\rangle.
\label{eq:thm2eq1}
\end{eqnarray}
Lemma~\ref{lem:sigma-spec} gives 
\begin{eqnarray}
Q_C^T|\tilde{v}\rangle\eqdef|\tilde{v}'\rangle&=&\frac{1}{\sqrt{2}}\left[
|\tilde{v}\rangle +\hat{C}|\tilde{v}\rangle\right]
=\frac{1}{\sqrt{2}}\left[|\tilde{v}\rangle+|\tilde{u}\rangle\right];
\label{eq:QConv}\\
Q_C^T|\tilde{u}\rangle\eqdef|\tilde{u}'\rangle&=&\frac{1}{\sqrt{2}}\left[
|\tilde{u}\rangle +\hat{C}|\tilde{u}\rangle\right]
=\frac{1}{\sqrt{2}}\left[|\tilde{u}\rangle +|\tilde{v}\rangle\right],\qquad
\label{eq:QConu}
\end{eqnarray}
where in the last line $\hat{C}|\tilde{u}\rangle=\hat{C}^2|\tilde{v}\rangle
=|\tilde{v}\rangle$ has been employed; note that $\hat{C}^2$ is the identity.
Equations~(\ref{eq:QConv}) and (\ref{eq:QConu}) reveal that 
$|\tilde{u}'\rangle=|\tilde{v}'\rangle$, so that Eq.~(\ref{eq:thm2eq1}) becomes
\begin{equation}
\langle u|U_{\rm HC}(t)|v\rangle=\langle\tilde{v}'|Q_C^TU_{\rm HC,I}(t)Q_C
|\tilde{v}'\rangle.
\label{eq:periodicityunderQC}
\end{equation}
This equation represents vertex periodicity in the adjacency matrix 
corresponding to the quotient graph under $\Pi_C$. If $t=t_p=\pi/2$ then using 
the results from Lemma~\ref{lem:sigma-spec}, Eq.~(\ref{eq:periodicityunderQC}) 
becomes
\begin{equation}
\langle u|U_{\rm HC}(t)|v\rangle=\gamma\langle\tilde{v}'|\tilde{v}'\rangle
=\gamma,
\end{equation}
where $\gamma=e^{-i\pi k(k-n)/2}$.

\end{proofthm}

\section{Discussion and Conclusions}
\label{sec:discussion}

In the preceding sections, we have shown that $k$ hard-core bosons on
hypercubically-weighted $n$-path graphs $\tilde{P}_n$ undergo perfect quantum
state transfer. The proofs of the main results, Theorem~\ref{thm:PST1} and 
Theorem~\ref{thm:PST2}, made use both of the celebrated Tonks-Girardeau 
solution (\ref{eq:TG}) for unweighted path graphs and of techniques from 
algebraic graph theory, such as Cartesian powers, graph isomorphisms, as well 
as equitable partitioning and its generalization to weighted graphs. The 
occurrence of PST was found to hinge on: (i) the 
preservation of the linear spectrum of the weighted path graph adjacency 
matrices in the graph for identical hard-core bosons; and (ii) the 
commutation of the unit antisymmetry operator ${\mathcal A}$ 
defined in Eq.~(\ref{eq:unit_adj}) with both the adjacency matrix of the 
Cartesian power graph (\ref{eq:construct-config}) and with the generalized 
mirror-symmetry operator $\hat{C}$, defined in Eq.~(\ref{eq:C-operator}). 
 
As discussed in Sec.~\ref{sec:interactions}, the graph for $k$ 
identical hard-core bosons corresponds to the $k$th symmetric power of 
the underlying graph~\cite{Audenaert2007}. This is obtained by initially taking 
the $k$th Cartesian power (representing $k$ distinguishable non-interacting 
bosons), followed by the deletion of all vertices representing 
multiple-occupancy of a given site. The graph is then equitably 
partitioned~\cite{AGT,BrouwerHaemers2012}, with vertices grouped according 
to their exchange symmetry, and the resulting graph quotient projects the 
system into the subspace of symmetric identical bosons. 

In contrast, 
the graph for $k$ indistinguishable non-interacting fermions corresponds to the 
exterior power of the underlying graph~\cite{Mallory2013}. This is obtained via 
the $k$th Cartesian power, followed by a signed projection into the subspace of 
antisymmetric identical fermions. 
Generally, the exterior power graph is not isomorphic to the symmetric power 
graph because of the presence of signed edges, even if their connectivity is 
identical. That said, if the exterior power graph is balanced then it is 
switching-equivalent to an all-positive graph~\cite{Mallory2013}; i.e.\ there 
exists a diagonal matrix with elements $\{+1,-1\}$, which is therefore also 
unitary, that transforms the exterior power graph to an unsigned graph. This 
work found that the only unsigned graphs on $n$ vertices that yield balanced 
exterior $k$th powers are path graphs for $k>1$ and $k<n$ or cycle graphs when 
$k$ is odd. In fact, for the path graph all edges of the exterior power are 
positive, so that no switching is required. The implication is that the graph 
for $k$ identical fermions on (weighted or simple) path graphs should be 
isomorphic to that for $k$ hard-core bosons, and therefore the two systems are 
wholly equivalent. Indeed, Corollary~\ref{cor:states} reveals that their 
spectra coincide.

What then is the essential difference between non-interacting fermions and 
hard-core bosons on path graphs that necessitates the TG solution in the first 
place? The answer is in the construction of the eigenstates themselves. The TG 
eigenfunctions~(\ref{eq:TG}) correspond to non-interacting 
{\it distinguishable} fermions; the number of the states is clearly $n^k$
rather than $n\choose k$. The coefficients of the fermionic states are derived 
from the single-particle eigenstates of the underlying path graph and are 
generally signed, even for the ground state (maximal eigenvector). Applying the 
indistinguishability equitable partitioning operator $Q_I^T$ would map all 
these states to null vectors because of the alternating signs. The unit 
antisymmetry operator ${\mathcal A}$, which symmetrizes the states, is 
therefore essential in order to obtain an explicit representation of the 
identical hard-core boson eigenstates. 

Alternatively, one can consider that the adjacency
matrix for $k$ distinguishable fermions on $P_n$ (or $\tilde{P}_n$) should not 
be considered to be that for the Cartesian power $A_{P_n^{\Box k}}$; rather, it
should be a signed graph with the same connectivity but where each vertex is 
switched (in the sense of Ref.~\cite{Mallory2013}) according to the 
antisymmetric ordering of its label $\ell_i$. The quotient of this graph under 
the equitable partition $\Pi_I$ would again be the null graph. The presence of 
${\mathcal A}$ is therefore crucial, and it is fortunate that 
$[{\mathcal A},\hat{C}]=0$ so that PST on weighted path graphs persists for 
hard-core bosons as it does for non-interacting fermions. In short, 
${\mathcal A}$ in the TG theory is the switching operator that maps the signed 
(but balanced) graph for non-interacting fermions to that for distinguishable
unsigned hard-core bosons. This interpretation also helps explain why the 
TG solution is specific to path and cycle graphs: no other graphs yield 
balanced antisymmetrized Cartesian powers.

This raises the important question: do hard-core bosons exhibit PST on more 
general graphs? The above discussion rules out a general mapping between
hard-core bosons and non-interacting fermions. The proofs of 
Theorems~\ref{thm:PST1} and \ref{thm:PST2} hinge on 
$[{\mathcal A},DA_{P_n^{\Box k}}D]=0$, which is no longer the case if the
vertex-deleted Cartesian-power adjacency matrix can no longer be expressed in
terms of $k!$ disconnected graph components. Furthermore, for more general 
graphs it is known that the ground state energy (maximal eigenvalue of the 
associated adjacency matrix) for hard-core bosons is always lower (higher) than 
that for non-interacting fermions~\cite{Nie2013}. Likewise, other 
eigenvalues are not likely to have any relationship with those of 
non-interacting fermions; satisfying Eq.~(\ref{eq:ratio}) is therefore 
unlikely. While one might be able to find graphs which exhibit PST for some 
specific number of particles, we conjecture that very few graphs (but most 
likely no graphs other than the weighted paths described here) will support PST 
for arbitrary numbers of hard-core bosons, even if they support PST for single 
particles. Unfortunately a proof of this conjecture is beyond the scope of the
current work.

\acknowledgments

The authors are grateful for insightful conversations with C.~Godsil and 
C.~Tamon. This work was supported by the Natural Sciences and Engineering 
Research Council of Canada (NSERC) and Alberta Innovates, Technology Futures 
(AITF).

\appendix

\section{Equitable partitioning of weighted graphs}
\label{equitableproof}

For a partition of a graph $\mathcal{G}$ to be equitable, the following
theorem must be satisfied.
\begin{theorem}
	\label{thm:eqptMain}
	Suppose $\Pi=\{C_i\}_{i=1}^\ncell$
	is a partition into $\ncell$ disjoint cells
	of the vertices of a connected undirected weighted
	graph $G$, and $Q$ is its
	normalized partition matrix as defined by Eq.~(\ref{eq:Qweighted}).
	Let $A$ be the adjacency matrix of $G$.
	Then the following are equivalent:
	\begin{enumerate}
		\item $\Pi$ is equitable. \label{itm:eqptThmTFAEequitable}
		\item The column space of $Q$ is $A$-invariant.
			\label{itm:eqptThmTFAEcolSpace}
		\item $A$ and $QQ^T$ commute. \label{itm:eqptThmTFAEcommute}
		\item There is an $\ncell\times\ncell$
			matrix $B$ such that $AQ=QB$. \label{itm:eqptThmTFAEexistsB}
	\end{enumerate}
\end{theorem}
Theorem~\ref{thm:eqptMain}
provides the main result that extends the notion
of equitable partitioning to weighted graphs. Its proof makes use of the
following lemma.

\begin{lemma}
	\label{lem:eqptQTQisI}
	Let $Q$ be the normalized partition matrix describing a partition
	$\Pi=\{C_i\}_{i=1}^\ncell$ of a weighted undirected connected graph
	$G$ on $\nvert$ vertices.
	Then $Q^TQ=I_\ncell$, the $\ncell\times\ncell$ identity.
\end{lemma}
\begin{prooflem}
	By the definition of $Q$, Eq.~(\ref{eq:Qweighted}),
	\begin{equation}
		Q^TQ
		=
		\sum_{v=1}^\nvert\sum_{w=1}^\nvert
			\Omega(v)\Omega(w)
			\ket{\idxof{v}}\!\!\braket{v}{w}\!\!\bra{\idxof{w}}
		=
		\sum_{v=1}^\nvert
			\Omega^2(v)\ketbra{\idxof{v}}{\idxof{v}}.
	\end{equation}
	This is a sum of $\nvert$ elements from a set of
	$\ncell$ distinct matrices, with $\nvert\ge\ncell$, so
	some of them have contributions from multiple vertex indices.
	Specifically, if two vertices $u$ and $v$ belong to the
	same cell then $\idxof{u}=\idxof{v}$, so it is natural to sum only
	over the cells, instead of over all vertices. Doing so and inserting
	Eq.~(\ref{eq:omega}), the definition of $\Omega$, yields
	\begin{eqnarray}
		Q^TQ
		&=&
		\sum_{i=1}^\ncell\sum_{v\in C_i}
		\frac{\omega^2(v)}{\omega^2(C_{\idxof{v}})}	
			\ketbra{\idxof{v}}{\idxof{v}}\nonumber \\
		&=&
		\sum_{i=1}^\ncell\frac{1}{\omega^2(C_i)}
		\left(\sum_{v\in C_i}
			\omega^2(v)
		\right)
		\ketbra{i}{i}.
	\end{eqnarray}
	The parenthetic term is simply the square of the weight of cell
	$C_i$ by definition, leaving
	\begin{equation}
		Q^TQ
		=
		\sum_{i=1}^\ncell
		\ketbra{i}{i}
		=
		I_\ncell,
	\end{equation}
	which completes the proof.
\end{prooflem}
Lemma 8.1 of Reference~\cite{Godsil-main} provides three statements related to
$A$, $Q$, and $\pi$, that are
mutually equivalent, as well as equivalent to the statement that $\Pi$
is equitable, in the case of simple graphs.
The same set of four statements can be shown to be equivalent
when $A$ describes a weighted graph, and the equitableness of a partition
on a weighted graph is defined as follows.
\begin{definition}\label{def:eqptWeightedEquitable}
A partition $\Pi$ of a weighted undirected connected graph $G$ is 
\emph{equitable} if,
for any pair of cells $C_i,C_j\in\Pi$, including $j=i$, and for
any vertex $u\in C_i$, the quantity
\begin{equation}
	b_{ij}(u)
	\eqdef
	\sum_{v\in C_j} A_{uv}\frac{\omega(v)}{\omega(u)}
\end{equation}
is a constant $b_{ij}$, independent of the choice of $u$.
\end{definition}
The above definition is justified by the following theorem, in which a set
of vectors is said to be $A$-invariant if a matrix $A$ maps each of the
vectors to a linear combination of one or more of them.
\begin{proofthm}
	The proof consists in showing that each statement implies its
	successor, cyclically.

(\ref{itm:eqptThmTFAEequitable}$\implies$\ref{itm:eqptThmTFAEcolSpace})
	To show that \ref{itm:eqptThmTFAEequitable} implies
	\ref{itm:eqptThmTFAEcolSpace}, suppose that $\Pi$ is equitable
	and consider column $i$ of $Q$, denoted
	\begin{equation}
		\ket{\phi_i}
		\eqdef
		\sum_{v\in C_i} \Omega(v) \ket{v}.
		\label{eq:eqptPhi_i}
	\end{equation}
	The action of $A$ on this column vector is
	\begin{align}
		A\ket{\phi_i}
		&=
		\sum_{v\in C_i}
			\Omega(v) A\ket{v}
		=
		\sum_{v\in C_i}
			\Omega(v)\sum_{u=1}^\nvert A_{uv}\ket{u} \nonumber\\
		&=
		\frac{1}{\omega(C_i)}
		\sum_{u=1}^\nvert
		\sum_{v\in C_i}\omega(v) A_{uv}\ket{u},
	\end{align}
	where the final equality follows from the definition
	of $\Omega(v)$, Eq.~(\ref{eq:omega}), 
	and the fact that the denominator is constant over the
	cell $C_i$.
	Since $\Pi$ is equitable by assumption, there exist constants
	$b_{\idxof{u}i}$ such that
	\begin{equation}
		\sum_{v\in C_i}\omega(v) A_{uv}
		=
		b_{\idxof{u}i}\omega(u),
	\end{equation}
	so the string of equalities continues as
	\begin{equation}
		A\ket{\phi_i}
		=
		\frac{1}{\omega(C_i)}
		\sum_{u=1}^\nvert b_{\idxof{u}i}\omega(u) \ket{u}
		=
		\sum_{j=1}^\ncell \frac{b_{ji}}{\omega(C_i)}
		\sum_{u\in C_j}\omega(u)\ket{u}.
	\end{equation}
	Finally, defining
	\begin{equation}
		{a}_{ij}
		\eqdef
		b_{ji}\frac{\omega(C_j)}{\omega(C_i)}
	\end{equation}
	leads to
	\begin{equation}
		A\ket{\phi_i}
		=
		\sum_{j=1}^\ncell a_{ij}
		\sum_{u\in C_j}\Omega(u)\ket{u}
		=
		\sum_{j=1}^\ncell a_{ij} \ket{\phi_j}.
	\end{equation}
	That is, $A$ takes each column of $Q$ to a linear combination of the
	columns of $Q$; the column space of $Q$ is $A$-invariant.

	({\ref{itm:eqptThmTFAEcolSpace}}$\implies${\ref{itm:eqptThmTFAEcommute}})
	Now assume that the column space of $Q$ is $A$-invariant. 
	With the definition (\ref{eq:eqptPhi_i}) of the $i$th column of $Q$,
	the partition matrix can be rewritten as
	\begin{equation}\label{eq:eqptQasPhis}
		Q
		=
		\sum_{i=1}^\ncell
			\ketbra{\phi_i}{i},
	\end{equation}
	and therefore
	\begin{equation}
		QQ^T
		=
		\sum_{i=1}^\ncell
			\ketbra{\phi_i}{i}
		\sum_{j=1}^\ncell
			\ketbra{j}{\phi_j}
		=
		\sum_{i=1}^\ncell
			\ketbra{\phi_i}{\phi_i}.
	\end{equation}
	The column space of $Q$ is $A$-invariant by assumption so for each
	$i\in\{1,\ldots,\ncell\}$,
	\begin{equation}\label{eq:eqptAonPhi}
		A\ket{\phi_i}
		=
		\sum_{j=1}^\ncell a_{ij}\ket{\phi_j}
		\quad\Rightarrow\quad
		\bra{\phi_i}A
		=
		\sum_{j=1}^\ncell a_{ij}\bra{\phi_j},
	\end{equation}
	since $A$ is real and symmetric, because $\mathcal{G}$ is assumed
	to be undirected.
	This is a non-trivial statement even though the $\ket{\phi_i}$ form an
	orthonormal set, since there are only $\ncell$ of them yet they are
	$\nvert$-component vectors, and therefore do not form a basis for
	$\mathbb{R}^\nvert$.
	The commutator of $A$ and $QQ^T$ can be written as
	\begin{align}
		AQQ^T-QQ^TA
		&=
		\sum_{i=1}^\ncell
			\sum_{j=1}^\ncell\left[ 
				a_{ij}\ketbra{\phi_j}{\phi_i}
				-
				a_{ij}\ketbra{\phi_i}{\phi_j}
			\right]\nonumber\\
		&=
		\sum_{i=1}^\ncell
			\sum_{j=1}^\ncell\left(
				a_{ij}-a_{ji}
			\right)\ketbra{\phi_i}{\phi_j},
	\end{align}
	but since the elements of $A$ and the $\ket{\phi_i}$ are real,
	and $A$ is symmetric,
	\begin{equation}\label{eq:eqptAijIsAji}
		a_{ij}
		=
		\bra{\phi_j}A\ket{\phi_i}
		=
		\bra{\phi_i}A\ket{\phi_j}
		=
		a_{ji}.
	\end{equation}
	Therefore the commutator vanishes, as required.

	({\ref{itm:eqptThmTFAEcommute}}$\implies${\ref{itm:eqptThmTFAEexistsB}})
	To prove that {\ref{itm:eqptThmTFAEcommute}} implies
	{\ref{itm:eqptThmTFAEexistsB}},
	assume now that $[A,QQ^T]=0$ and consider the matrix
	$\tilde{B}=Q^TAQ$, from which one sees
	\begin{equation}
		Q\tilde{B}
		=
		QQ^TAQ
		=
		AQQ^TQ.
	\end{equation}
	But Lemma~\ref{lem:eqptQTQisI} shows that $Q^TQ$ is the identity,
	leaving $Q\tilde{B}=AQ$.
	Therefore, $\tilde{B}$ is in fact the required $\ncell\times\ncell$ matrix
	$B$.

	({\ref{itm:eqptThmTFAEexistsB}}$\implies${\ref{itm:eqptThmTFAEequitable}})
	Finally, to see that {\ref{itm:eqptThmTFAEexistsB}} implies
	{\ref{itm:eqptThmTFAEequitable}}, which will complete the proof,
	suppose there exists an $\ncell\times\ncell$ matrix $B$ such that $QB=AQ$.
	Then by definition,
	\begin{equation}
		\sum_{v=1}^\nvert\Omega(v)\ketbra{v}{\idxof{v}}B
		=
		\sum_{v=1}^\nvert\Omega(v)A\ketbra{v}{\idxof{v}}.
	\end{equation}
	Multiplying each side by $\bra{u}$ on the left and $\ket{j}$ on the
	right leads to
	\begin{equation}
		\Omega(u)B_{\idxof{u}j}
		=
		\sum_{v=1}^\ncell\Omega(v)A_{uv}\delta_{\idxof{v}j}
		=
		\sum_{v\in C_j}\Omega(v)A_{uv}.
	\end{equation}
	From the definition of $\Omega$	one obtains
	\begin{equation}
		\frac{\omega(u)}{\omega(C_{\idxof{u}})}
		B_{\idxof{u}j}
		=
		\frac{1}{\omega(C_j)}\sum_{v\in C_j}\omega(v) A_{uv},
	\end{equation}
	which, with $i=\idxof{u}$, in turn implies that
	\begin{equation}
		\sum_{v\in C_j} A_{uv}\frac{\omega(v)}{\omega(u)}
		=
		B_{ij}\frac{\omega(C_j)}{\omega(C_{i})}
		\eqdef
		b_{ij},
	\end{equation}
	from which the conclusion immediately follows since the final
	right-hand
	side is independent of the choice of $u$ from $C_{i}$, and
	thus $Q$ encodes an equitable partition, $\Pi$.
\end{proofthm}
Statement {\ref{itm:eqptThmTFAEcommute}} of Theorem
\ref{thm:eqptMain}, that $[A,QQ^T]=0$, leads to the following
useful result.
\begin{cor}\label{cor:eqptQQTEvals}
	Suppose $\Pi$ is an equitable partition of a connected, weighted, undirected 
	graph $\mathcal{G}$. Let $A$ be the adjacency matrix of $\mathcal{G}$,
	and $Q$ the normalized partition matrix of $\Pi$.
	Then $QQ^T$ has two distinct eigenvalues, $0$ and $1$.
\end{cor}
\begin{proofcor}
	Since $\Pi$ is equitable, Theorem~\ref{thm:eqptMain} shows that
	$A$ and $QQ^T$ commute. Both are real and symmetric, and therefore
	diagonalizable, thus they are simultaneously diagonalizable and
	share an eigenspace.
	Let the eigenvalues of $A$ be $\lambda_i$, with corresponding
	eigenvectors $\ket{\lambda_{i,j}}$ where for each $i$, $j$ runs from
	$1$ to the multiplicity of $\lambda_i$.
	Let $q_{i,j}$ be the eigenvalues of $QQ^T$, so that
	\begin{equation}
		QQ^T\ket{\lambda_{i,j}}
		=
		q_{i,j}\ket{\lambda_{i,j}}.
	\end{equation}
	Left-multiplying this expression by $Q^T$ yields
	$Q^T\ket{\lambda_{i,j}}=q_{i,j}Q^T\ket{\lambda_{i,j}}$,
	from which it can be concluded that either $q_{i,j}=1$ or
	$Q^T\ket{\lambda_{i,j}}=0$.
	Clearly if the latter is the case, then it is also true that
	$QQ^T\ket{\lambda_{i,j}}=0$ which must still equal
	$q_{i,j}\ket{\lambda_{i,j}}$. But the $\ket{\lambda_{i,j}}$ are
	non-zero vectors, so in this case $q_{i,j}$ must vanish.
	Therefore either $q_{i,j}=1$ or $q_{i,j}=0$.
\end{proofcor}

Given a weighted, connected, undirected 
graph $\mathcal{G}$ and an equitable partition $\Pi$ of its vertices
into $\ncell$ cells, with adjacency and normalized partition matrices
$A$ and $Q$ respectively,
Theorem~\ref{thm:eqptMain} guarantees that there exists an
$\ncell\times\ncell$ matrix $B$ such that $AQ=QB$.
Lemma~\ref{lem:eqptQTQisI} can then be used to show that
\begin{equation}
	B = Q^TAQ.
\end{equation}
$B$ is also a real, symmetric matrix,
since
\begin{equation}
	\bra{i}B\ket{j}
	=
	\bra{\phi_i}A\ket{\phi_j}
	=
	\bra{\phi_i}A^T\ket{\phi_j}
	=
	\bra{\phi_j}A\ket{\phi_i}
	=
	\bra{j}B\ket{i}.
\end{equation}
Therefore, $B$ can be interpreted as the adjacency matrix of a graph on
$\ncell$ vertices; this graph is called the quotient graph
of $\mathcal{G}$ with respect to $\Pi$ and is denoted $\mathcal{G}/\Pi$.

\section{Eigenvalues of graphs under equitable partitioning}
\label{sec:EquitablePartitioning}

A remarkable property of the quotient graph is that it shares all of its 
eigenvalues with the original one. The following lemma makes this statement 
concrete, and is subsequently used to prove that for all non-negative graphs 
(including the multi-boson graphs considered in the present work) the maximal 
eigenvalue is always preserved in the quotient. 

\begin{lemma}
	If $B=Q^TAQ$ is the adjacency matrix of a quotient graph $\mathcal{G}/\Pi$ 
	under an equitable partition $\Pi$ from a weighted, undirected, connected 
	graph $\mathcal{G}$, then every eigenvalue of $B$ is also an eigenvalue of 
	$A$.
	\label{lem:eqptEvalSubset}
\end{lemma}

\begin{prooflem}
Let $\beta$ be an eigenvalue of $B$, with corresponding
eigenvector $\ket{\beta}$. 
Since $\Pi$ is equitable, $AQ=QB$ by Theorem~\ref{thm:eqptMain}.
Therefore
\begin{equation}
AQ\ket{\beta}=QB\ket{\beta}=\beta Q\ket{\beta},
\end{equation}
and $\beta$ is seen to be an eigenvalue of $A$, as required.
\end{prooflem}
The proof of Lemma~\ref{lem:eqptEvalSubset}
additionally shows that the eigenvectors $\ket{\beta}$ of $B$ are
related to a subset of the eigenvectors $\ket{\alpha}$ of $A$ by
$\ket{\beta}=Q^T\ket{\alpha}$ and $\ket{\alpha}=Q\ket{\beta}$.
The determination of which eigenvectors belong to this subset in general
remains an open question, though Theorem~\ref{thm:eqptMaxEvalPreserved}
below shows that the eigenvector corresponding to the maximal eigenvalue is
preserved under collapse whenever every entry of $A$ is non-negative in
the vertex basis.

The following lemma shows that not only does every eigenvector of $B$ yield an
eigenvector of $A$ under the action of $Q$, but also that every eigenvector of 
$A$ that does not vanish under the action of $Q^T$ yields an eigenvector of
$B$. 
\begin{lemma}
	\label{lem:eqptAVecsToB}
	Let $\mathcal{G}$ be a weighted, undirected, connected graph with adjacency
	matrix $A$, and $\Pi$ an equitable partition with normalized
	partition matrix $Q$ that generates the
	quotient graph $\mathcal{G}/\Pi$ with adjacency matrix $B=Q^TAQ$.
	Suppose $\ket{\lambda_{i,j}}$ is an eigenvector of $A$ as defined in
	the proof of Corollary~\ref{cor:eqptQQTEvals}.
	Then either $Q^T\ket{\lambda_{i,j}}=0$ or $Q^T\ket{\lambda_{i,j}}$
	is an eigenvector of $B$.
\end{lemma}
\begin{prooflem}
	Corollary~\ref{cor:eqptQQTEvals} shows that the eigenvalues of $QQ^T$
	associated with the eigenvectors $\ket{\lambda_{i,j}}$ are
	$q_{i,j}\in\{0,1\}$. Suppose that $QQ^T\ket{\lambda_{i,j}}=0$.
	Then since $Q^TQ=I$, left-multiplication by $Q^T$ yields
	$Q^T\ket{\lambda_{i,j}}=0$. On the other hand, in the case
	$QQ^T\ket{\lambda_{i,j}}=\ket{\lambda_{i,j}}$, one obtains
	\begin{equation}
		A\left(QQ^T\ket{\lambda_{i,j}}\right)
		=
		A\ket{\lambda_{i,j}}
		=
		\lambda_i\ket{\lambda_{i,j}}.
	\end{equation}
	Left-multiplying the initial and final expressions by $Q^T$ 
	yields
	\begin{equation}
		(Q^T AQ)Q^T\ket{\lambda_{i,j}}
		=
		B\left(Q^T\ket{\lambda_{i,j}}\right)
		=
		\lambda_i\left(Q^T\ket{\lambda_{i,j}}\right),
	\end{equation}
	showing that $Q^T\ket{\lambda_{i,j}}$ is an eigenvector of $B$,
	with eigenvalue $\lambda_i$.
\end{prooflem}

Some further definitions and results from the fields of linear algebra and 
graph theory that are useful during the proof of the upcoming 
Theorem~\ref{thm:eqptMaxEvalPreserved} are stated here without proof. A 
treatment can be found, for example, in Ref.~\cite{BrouwerHaemers2012}.
\begin{definition}
An $\nvert\times\nvert$ matrix $T$ is \emph{irreducible} if for each
$i,j\in\{1,\ldots,\nvert\}$ there exists a positive integer $k$ such that
$(T^k)_{ij}>0$. A graph $\mathcal{G}$ is said to be
irreducible if its
adjacency matrix $A(\mathcal{G})$ is irreducible.
\end{definition}
The following lemma relates the irreducibility of
a matrix to connectedness of a corresponding unweighted graph.
\begin{lemma}\label{lem:eqptIrreducible}
	An $\nvert\times\nvert$ matrix $T$ with elements $T_{ij}$ is
	irreducible 
	if and only if the unweighted directed graph $\Gamma_T$, defined
	on the vertex set $\{1,\ldots,\nvert\}$ with an edge from $i$ to $j$
	whenever $T_{ij}>0$, is strongly connected. 
\end{lemma}
A graph is strongly connected
if there is a (directed) path from each vertex to every other vertex.
Therefore an undirected graph is strongly connected if and only if it is
connected.

Another useful result is the Perron-Frobenius theorem.


\begin{theorem}[Perron-Frobenius]
\label{thm:PerronFrobenius}
{\it 
	\label{thm:PF}
	Let $T$ be a non-negative irreducible square matrix.
	Then there exists a real number $\lambda_1>0$ with the
	following properties:
	\begin{enumerate}
		\item There exists a real vector $\ket{\lambda_1}$
			with each entry strictly positive, and such that
			$T\ket{\lambda_1}=\lambda_1\ket{\lambda_1}$.
		\item The algebraic and geometric multiplicities of
			$\lambda_1$ are both equal to 1. That is, 
			its associated eigenspace is one-dimensional.
		\item For each eigenvalue $\lambda_i$ of $T$, 
			$|\lambda_i|\le\lambda_1$.
	\end{enumerate}
}
\end{theorem}
This completes the prerequisites for the proof of the following theorem.

\begin{theorem}
	Let $\mathcal{G}$ be a connected undirected graph with
	non-negative weights and adjacency matrix $A$, and let $\Pi$
	be an equitable partition of its vertices with corresponding
	normalized partition matrix $Q$. Then the largest eigenvalue of
	$A$ is unique, and equal to the unique largest
	eigenvalue of the collapsed graph with adjacency matrix
	$B=Q^TAQ$.
	\label{thm:eqptMaxEvalPreserved}
\end{theorem}
\begin{proofthm}
	Let the eigenvalues of $A$ be $\lambda_i$, with corresponding
	eigenvectors $\ket{\lambda_{i,j}}$, where for each $i$,
	$j$ runs from 1
	to the algebraic multiplicity of $\lambda_i$.
	According to Lemma~\ref{lem:eqptEvalSubset}, each eigenvector of
	$B$ is given by $Q^T\ket{\lambda_{i,j}}$ for some $i$ and $j$.
	The eigenvalues of $B$ are therefore those $\lambda_i$ for
	which there exists at least one $j$ such that
	$Q^T\ket{\lambda_{i,j}}\neq 0$.
	Consider then,
	\begin{eqnarray}
		Q^T\ket{\lambda_{i,j}}
		&=&
		\sum_{v=1}^\nvert \Omega(v)\ket{\idxof{v}}\!\!
		\braket{v}{\lambda_{i,j}}\nonumber \\
		&=&
		\sum_{k=1}^\ncell\left(
			\sum_{v\in C_k}\Omega(v)\braket{v}{\lambda_{i,j}}
		\right)\ket{k},\label{eq:eqptQTlambda_by_k}
	\end{eqnarray}
	which shows that $Q^T\ket{\lambda_{i,j}}$ is non-zero
	if there exists at least one $k$ for which the parenthetic
	coefficient does not vanish.

	Since $\mathcal{G}$ is assumed to be connected and undirected, it is strongly
	connected; it contains no negative edge weights, satisfying
	the requirements of Lemma~\ref{lem:eqptIrreducible} and the
	Perron-Frobenius theorem.
	Therefore there is a unique largest eigenvalue
	$\lambda_1$ of $A$, with a single corresponding eigenvector
	$\ket{\lambda_{1,1}}$. The elements $\braket{v}{\lambda_{i,j}}$
	of this vector are strictly positive, and by definition
	$\Omega(v)>0$ for any vertex with at least one neighbour.
	Since $\mathcal{G}$ is connected, it contains no isolated
	vertices. As such, for every $k$ the coefficient in
	Equation~\eqref{eq:eqptQTlambda_by_k} is a sum of one or more
	strictly positive numbers
	and therefore cannot vanish.
	Thus $Q^T\ket{\lambda_{1,1}}\neq 0$.
	So by Lemma~\ref{lem:eqptAVecsToB},
	$Q^T\ket{\lambda_{1,1}}$ is an eigenvector of $B$, with eigenvalue
	$\lambda_1$.
	Furthermore, it can be seen from the proof of
	Lemma~\ref{lem:eqptAVecsToB} that $\lambda_1$ is the unique largest
	eigenvalue of $B$ as well as of $A$.
\end{proofthm}

\bibliography{hcpst}

\end{document}